\documentclass[12pt,a4paper]{article}
\usepackage{amssymb,bm}
\newcommand{\Nis}{n_c}
\newcommand{\tr}{{\mathrm{Tr}}}
\newcommand{\R}{\mathbb{R}}
\newcommand{\Z}{\mathbb{Z}}
\newcommand{\N}{\mathbb{N}}
\newcommand{\calF}{\mathcal{F}}
\newcommand{\calG}{\mathcal{G}}
\newcommand{\epn}{\epsilon}
\newcommand{\p}{\partial}
\newcommand{\ptau}{\partial_\tau}
\newcommand{\nn}{\nonumber\\}
\newcommand{\com}[1]{[\,#1\,]}
\newcommand{\acom}[1]{\{\,#1\,\}}
\newcommand{\scom}[1]{[\,#1\,]_*}
\newcommand{\gIIB}{g_b}
\newcommand{\tgIIB}{\tilde{g}_b}
\newcommand{\gym}{g_{Y\!M}}
\newcommand{\ogym}{\bar{g}_{Y\!M}}
\newcommand{\cpq}{c_{pq}}
\newcommand{\Opq}{O_{(p,q)}}
\newcommand{\hp}{\hat{p}}
\newcommand{\hq}{\hat{q}}
\newcommand{\hS}{\hat{\Sigma}}
\newcommand{\tg}{\tilde{g}}
\newcommand{\tA}{\tilde{A}}
\newcommand{\tG}{\tilde{G}}
\newcommand{\tR}{\tilde{R}}
\newcommand{\tX}{\tilde{X}}
\newcommand{\thS}{\tilde{\hat{\Sigma}}}
\newcommand{\tBNS}{\tilde{B}^{(1)}}
\newcommand{\tBR}{\tilde{B}^{(2)}}
\newcommand{\BNS}{B^{(1)}}
\newcommand{\BR}{B^{(2)}}

\newcommand{\volt}{V_{T^2}}
\newcommand{\Smr}{S_{M\!R}}

\title{($p,q$)-string in matrix-regularized membrane\\
and type IIB duality}
\author{
{\sc Hiroyuki Okagawa$^1$}\footnote{email:
	okagawa@eken.phys.nagoya-u.ac.jp},~
{\sc Shozo Uehara$^2$}\footnote{e-mail:
	uehara@is.utsunomiya-u.ac.jp}~ and
{\sc Satoshi Yamada$^1$}\footnote{e-mail:
	yamada@eken.phys.nagoya-u.ac.jp}\vspace{4mm}\\
{\sl $^1$Department of Physics, Nagoya University,}\\
{\sl Chikusa-ku, Nagoya 464-8602, Japan,}\\
{\sl $^2$Department of Information Science, Utsunomiya University,}\\
{\sl Utsunomiya 321-8585, Japan,}}
\date{}
\makeatletter

 \@addtoreset{equation}{section}
\renewcommand{\thefigure}{\@arabic\c@figure}
\makeatother
\begin{document}
\maketitle
\vspace{-255pt}
\begin{flushright}
	arXiv:0708.3484\\
	August 2007
\end{flushright}
\vspace{57mm}

\begin{abstract}
We consider a lightcone wrapped supermembrane compactified on a
2-torus in the matrix regularization.
We examine the double dimensional reduction technique and deduce the
free matrix string of ($p,q$)-string in type IIB superstring theory
explicitly from the matrix-regularized wrapped supermembrane.
In addition we obtain the (2+1)-dimensional super Yang-Mills
action in a curved background. We also examine the $SL(2)$ duality in
type IIB theory.
\end{abstract}

\newpage

\section{Introduction}
M-theory in eleven dimensions \cite{Wit,Tow} unifies all the five
perturbative superstring theories in ten dimensions and it should be
reduced to 11-dimensional supergravity theory in its low energy limit.
Supermembrane in eleven dimensions \cite{BST} is believed to play an
important role to understand the dynamics of M-theory.
In fact, it was shown that the wrapped supermembrane on
$\R^{10}\times S^1$ leads to the type IIA fundamental strings in the
shrinking limit of $S^1$, or by means of the double dimensional
reduction \cite{DHIS}.
On the other hand, type IIA superstring theory on $\R^9\times S^1$ is
equivalent via T-duality to the type IIB superstring theory on
$\R^9\times \tilde{S}^1$, where $\tilde{S}^1$ is the dual circle whose
radius is inversely related to the one of the $S^1$ \cite{DHS,DLP}.
And the shrinking $S^1$ limit of the type IIA superstring theory on
$\R^9\times S^1$ leads to the type IIB superstring theory in $\R^{10}$.
Accordingly, M-theory on $\R^9\times T^2$ in the shrinking volume
limit of $T^2$ is reduced to the type IIB superstring theory in
$\R^{10}$ \cite{Asp}.
It was shown that the type IIB superstring theory contains a bound
state of $p$ fundamental strings and $q$ D1-branes (D-strings), which
is called a $(p,q)$-string \cite{Sch,W}.
It was pointed out that the supermembrane wrapped $p$ times around a
compactified direction and $q$ times around the other compactified
direction of the target-space is reduced to a $(p,q)$-string
\cite{Sch}.
Recently, type IIB $(p,q)$-string action was deduced directly from the
wrapped supermembrane action on $\R^9\times T^2$ adopting the double
dimensional reduction and T-duality \cite{OUY}, in which the reduced
supermembrane is coupled to both the RR and NSNS 2-forms and it has
the correct tension of $(p,q)$-string.

Supermembrane theory is self-interacting and it has continuous energy
spectrum \cite{dWLN}. This implies that it is inherently multi-body
and has no coupling constant.
Thus we cannot directly adopt the ordinary canonical quantization
procedure to the supermembrane theory.
In order to handle the supermembrane the matrix regularization was
introduced as a ``quantization" procedure of the supermembrane
\cite{Hop,dWHN}.
Matrix theory \cite{BFSS} is described by $N\times N$ matrices which
can be thought of as the spatial component of 10-dimensional super
Yang-Mills fields after reducing to 1+0 dimension.
This supersymmetric quantum mechanical system is interpreted as
the low energy effective theory of D0-branes (D-particles).
And it was conjectured that the $N\to\infty$ limit of the system
captures all the degrees of freedom of M-theory in the infinite
momentum frame.

Matrix theory compactified on $S^1$ leads to matrix string theory
\cite{Mot,BS,DVV} through the T-duality prescription \cite{Tay}.
And hence matrix string theory can be thought of as (1+1)-dimensional
super Yang-Mills theory describing the low energy effective theory
of D-strings.
It is also conjectured to give a non-perturbative definition of the
type IIA superstring theory.
Similarly, the matrix regularization of wrapped supermembrane on
$\R^9\times S^1$ leads to matrix string theory \cite{SY,Ced,UY3}.
Furthermore, the matrix regularization procedure of wrapped
supermembrane on $\R^9 \times T^2$ was introduced \cite{UY4} and it
was shown that the regularized theory is T-dual to (2+1)-dimensional
super Yang-Mills theory \cite{UY4} which is low energy effective
theory of D2-branes.

The purpose of this paper is to deduce matrix $(p,q)$-strings directly
from a matrix-regularized lightcone supermembrane compactified on
a 2-torus referring to the analysis in ref.\cite{OUY}.
In addition, following the lines of \cite{UY4}, we will obtain the
(2+1)-dimensional super Yang-Mills theory in a curved background from
the matrix regularized wrapped supermembrane and we examine the
duality in the dimensionally reduced type II string. We should note
that the curved background fields are not mapped to matrix-valued
background fields, or they are proportional to the unit matrix in the
matrix regularization.
The background plays the role of probing the membrane or
($p,q$)-string. That is, we shall see that both the NSNS and RR
2-forms are coupled to the matrix regularized ($p,q$)-string.

The plan of this paper is as follows.
In section \ref{S:MS} we mainly review the matrix regularization of
the lightcone wrapped supermembrane on $\R^9\times T^2$ \cite{UY4} to
fix the notations, which are used in the following sections.
In section \ref{S:L} we will consider a lightcone wrapped
supermembrane compactified on a 2-torus in a curved background and
apply matrix regularization technique to it.
Then we adopt the double dimensional reduction technique and derive
the Green-Schwarz $(p,q)$-string.
In section \ref{S:YMPQ} we also start with the wrapped supermembrane.
Then we apply the matrix regularization technique with a suitable
choice of the matrix representation to give a standard form of super
Yang-Mills action in a curved background and we consider the
$SL(2,\R)$ transformation and type IIB string duality.
The section \ref{S:SD} is devoted to summary and discussion.

\section{Matrix-regularized wrapped supermembrane in flat
	background}\label{S:MS}
The 11-dimensional supermembrane in the lightcone gauge\footnote{
In this paper, we consider only toroidal membranes. Precisely
speaking, in this case, we need to impose the global constraints
associated with the information of the global topology\cite{UY2}.}
is given by (only bosonic degrees of freedom are presented here)
\begin{eqnarray}
  S&=&\frac{LT}{2}\int d\tau \int_0^{2\pi}\!d\sigma^1 d\sigma^2
    \left[(D_{\tau}X^M)^2 -\frac{1}{2L^2}\acom{X^M,X^N}^2\right]
	\label{eq:MLC},\\
  &&D_{\tau}X^M=\ptau X^M -\frac{1}{L}\acom{A,X^M},\\
  &&\acom{A,B}\equiv\epn^{ij} \p_{\sigma^i} A \p_{\sigma^j} B\,,
	\label{eq:PB}
\end{eqnarray}
where $\p_\tau =\p/\p \tau$, $\p_{\sigma^i}=\p/\p \sigma^i$,
$i,j=1,2$, $\epn^{12}=-\epn^{21}=1,\epn^{11}=\epn^{22}=0$,
$M,N=1,2,\cdots,9$, $X^M$ is the target-space coordinates and $A$ is
the gauge field, $T$ is the tension of the supermembrane and $L$ is an
arbitrary parameter of mass dimension $-1$.\footnote{The mass
dimensions of the world-volume parameters, $\tau,\sigma^1$ and
$\sigma^2$, are $0$.}
This theory has the area preserving diffeomorphisms (APD) of the
spacesheet as a residual symmetry.
Note that $L$ can be changed for $L'$ by a simple rescaling
of $\tau\to (L/L')\,\tau$.

Let us consider the wrapped supermembrane on $\R^9\times T^2$ taking
$X^8$ and $X^9$ as the coordinates of the two cycles of the $T^2$.
Then the target-space coordinates $X^M$ and the gauge $A$ are
expanded as\footnote{The $\tau$-dependence is not written explicitly
for all the variables.}
\begin{eqnarray}
 X^9(\sigma^1,\sigma^2) &=&w_1L_1\sigma^2 +\sum_{k_1,k_2
    =-\infty}^{\infty}Y^1_{(k_1,k_2)}\,e^{ik_1\sigma^1+ik_2\sigma^2}
	= w_1L_1\sigma^2 + Y^1(\sigma^1,\sigma^2),\label{eq:md1}\\
 X^8(\sigma^1,\sigma^2)&=&w_2L_2\sigma^1 +\sum_{k_1,k_2
    =-\infty}^{\infty}Y^2_{(k_1,k_2)}\,e^{ik_1\sigma^1+ik_2\sigma^2}
    \equiv w_2L_2\sigma^1 + Y^2(\sigma^1,\sigma^2),\label{eq:md2}\\
 X^m(\sigma^1,\sigma^2)&=&\sum_{k_1,k_2
     =-\infty}^{\infty}X^m_{(k_1,k_2)}\,
	e^{ik_1\sigma^1+ik_2\sigma^2},\label{eq:md3}\\
 A(\sigma^1,\sigma^2)&=&\sum_{k_1,k_2=-\infty}^{\infty}A_{(k_1,k_2)}\,
	e^{ik_1\sigma^1+ik_2\sigma^2},\label{eq:md4}
\end{eqnarray}
where $m=1,2,\cdots,7$, $L_1$ and $L_2$ are the radii of the two
cycles of $T^2$ and $w_1,w_2\,(\ne0)$ are integers.
These fields satisfy the periodicity conditions,
\begin{eqnarray}
 X^{9}(\sigma^1,\sigma^2+2\pi)&=&2\pi w_1 L_1
	+X^{9}(\sigma^1,\sigma^2)\,,\label{eq:01p}\\
 X^{8}(\sigma^1,\sigma^2+2\pi)&=&X^{8}(\sigma^1,\sigma^2)\,,
	\label{eq:01q}\\
 X^{9}(\sigma^1+2\pi,\sigma^2)&=&X^{9}(\sigma^1,\sigma^2)\,,
	\label{eq:02r}\\
 X^{8}(\sigma^1+2\pi,\sigma^2)&=&2\pi w_2 L_2
	+X^{8}(\sigma^1,\sigma^2)\,,\label{eq:02s}\\
 X^m(\sigma^1+2\pi,\sigma^2)&=&X^m(\sigma^1,\sigma^2+2\pi)
	=X^m(\sigma^1,\sigma^2)\,,\label{eq:02t}\\
 A(\sigma^1+2\pi,\sigma^2)&=&A(\sigma^1,\sigma^2+2\pi)
	=A(\sigma^1,\sigma^2)\,.\label{eq:02u}
\end{eqnarray}
These represent the supermembrane wrapping $w_1$-times around one of
the two compact directions $X^9$ and $w_2$-times around the other
direction $X^8$.
We call these two cycles $(0,1)$- and $(1,0)$-cycles, respectively.
Plugging eqs.(\ref{eq:md1})--(\ref{eq:md4}) into the covariant
derivatives and the Poisson brackets, we have
\begin{eqnarray}
 F_{\tau\sigma^1}&\equiv&D_{\tau}X^9=\p_{\tau}Y^1
	-\frac{w_1L_1}{L}\p_{\sigma^1}A
	-\frac{1}{ L}\acom{A,Y^1}\,,\label{eq:dt9}\\
 F_{\tau\sigma^2}&\equiv&D_{\tau}X^8=\p_{\tau} Y^2
	+\frac{w_2L_2}{L}\p_{\sigma^2}A
	-\frac{1}{L}\acom{A, Y^2}\,,\\
 F_{\sigma^1\sigma^2}&\equiv&\frac{1}{L}\,\acom{X^8,X^9}=
	\frac{w_1w_2L_1L_2}{L}+\frac{w_1L_1}{L}\,\p_{\sigma^1}Y^2\nn
  &&\phantom{\frac{1}{L}\,\acom{X^8,X^9}=\quad}
	+\frac{w_2L_2}{L}\,\p_{\sigma^2}Y^1
	+\frac{1}{L}\,\acom{Y^2,Y^1}\,,\\
 D_{\tau}X^m&=&\p_{\tau} X^m -\frac{1}{L}\,\acom{A,X^m}\,,\\
 D_{\sigma^1} X^m&\equiv &\frac{1}{L}\,\acom{X^9,X^m}=
	-\frac{w_1L_1}{L}\,\p_{\sigma^1} X^m
	+\frac{1}{L}\,\acom{Y^1,X^m}\,,\\
 D_{\sigma^2} X^m&\equiv &\frac{1}{L}\,\acom{X^8,X^m}
	=\frac{w_2L_2}{L}\,\p_{\sigma^2} X^m
	+\frac{1}{L}\,\acom{Y^2,X^m}\,.\label{eq:drx}
\end{eqnarray}
Thus, the action (\ref{eq:MLC}) is rewritten by
\begin{eqnarray}
  S&=&\frac{LT}{2}\int d\tau \int_0^{2\pi}\!d\sigma^1 d\sigma^2
    \Biggl[{F}_{\tau \sigma^1 }^2+{F}_{\tau \sigma^2 }^2
       -{F}_{\sigma^1 \sigma^2}^2 \nn
  &&\qquad +({D}_\tau X^m)^2 -({D}_{\sigma^1} X^m )^2
    -({D}_{\sigma^2} X^m)^2 -\frac{1}{2L^2}\acom{X^m,X^n}^2\Biggr]\,.
	\label{eq:MLC0}
\end{eqnarray}

\subsection{The matrix representation}\label{S:MR}
Here we shall consider the matrix regularization of the wrapped
supermembrane on $\R^9 \times T^2$, eqs.(\ref{eq:MLC})-(\ref{eq:md4}).
The procedure for the matrix regularization is the following
\cite{UY4}: (i) Introduce the noncommutativity on the
spacesheet of supermembrane, or replace the product of
functions on the spacesheet to the star-product.
(ii) If possible, find the central elements of the star-commutator
algebra and truncate the generators of the algebra consistently.
(iii) Give a matrix representation of the (truncated) star-commutator
algebra.

The star-commutators algebra for the set of generators
$\{e^{ik_1\sigma^1+ik_2\sigma^2},\sigma^1,
\sigma^2\,|\,k_1,k_2\in\Z\}$ is given by \cite{Ced,UY3,UY4,FZ}
\begin{eqnarray}
  \scom{e^{ik_1\sigma^1+ik_2\sigma^2},e^{ik'_1\sigma^1+ik'_2\sigma^2}}
  &=&-2i\sin\left(\frac{\pi}{N}\,\epn^{ij} k_ik_j'\right)\,
	e^{i(k_1+k'_1)\sigma^1+i(k_2+k'_2)\sigma^2}\,,\label{eq:al1}\\
  \scom{\sigma^1,e^{ik_1\sigma^1+ik_2\sigma^2}}&=&
    -\frac{2\pi k_2}{N}\,
	e^{ik_1\sigma^1+ik_2\sigma^2}\,,\label{eq:al2}\\
  \scom{\sigma^2,e^{ik_1\sigma^1+ik_2\sigma^2}}&=&
    \frac{2\pi k_1}{N}\,
	e^{ik_1\sigma^1+ik_2\sigma^2}\,,\label{eq:al3}\\
  \scom{\sigma^1,\sigma^2}&=&i\frac{2\pi}{N}\,.\label{eq:al4}
\end{eqnarray}
Since we can not find a central element of the algebra
eqs.(\ref{eq:al1})-(\ref{eq:al4}), the truncation is not possible in
$T^2$ compactified case.
The generators are represented by $N\times N$ matrices with two
continuous parameters $\theta^1, \theta^2$ \cite{UY4},\footnote{Note
that the parameters $\theta^i$ are, in principle, independent of the
spacesheet coordinates $\sigma^1,\sigma^2$.}
\begin{eqnarray}
 e^{i(u_1N+v_1)\sigma^1+i(u_2N+v_2)\sigma^2} &\to&
    e^{i(u_1N+v_1)\theta^1/N}e^{-i(u_2N+v_2)\theta^2/N}
    \lambda^{-v_1v_2/2}\,V^{v_2}\,U^{v_1},\label{eq:mx1}\\
 \sigma^2&\to&-2\pi i \p_{\theta^1} I_N,\label{eq:mx2}\\
 \sigma^1&\to&-2\pi i \p_{\theta^2}I_N
	+\frac{\theta^1}{N}\,I_N\,,\label{eq:mx3}
\end{eqnarray}
where $\lambda = e^{i2\pi/N}$, $u_1,u_2\in\Z$,
$v_1,v_2=0,\pm1,\pm2,\cdots, \pm M,$\footnote{We assume $N$ odd,
$N=2M+1$ and we have parametrized $k_i$ as $k_i=u_i
N+v_i\ (i=1,2)$.} $I_N$ is the $N\times N$ unit matrix and $U, V$ are
the $N\times N$ clock and shift matrices, respectively,
\begin{eqnarray}
  U = \left(\begin{array}{ccccc}
	1 & & & &  \lower10pt\hbox{\Large 0}\\[-5pt]
	 & \lambda& & &  \\
	 &  & \lambda^2& &  \\
	 &  &  & \ddots &\\
	&\hbox{\Large0} & & &\lambda^{N-1}\end{array}\right),
	\label{eq:U}\ \quad
 V = \left(\begin{array}{ccc@{}c@{}cc}
	0& 1&  & & & \\
	 & 0& 1& & & \\
	\vdots &  &  &\ddots&\ddots  & \\
	0&  &  & & 0& 1\\
	1& 0&  &\cdots & &0\end{array}\right) .
\end{eqnarray}
These have the following properties,
\begin{equation}
	\quad U^N=V^N=I_N\,,\quad VU=\lambda\,UV\,.
\end{equation}
Then, the functions $X^9,X^8,X^m$ and $A$ of $\sigma^1$ and
$\sigma^2$,\footnote{Of course they are also functions of $\tau$.
We just do not mention it explicitly.}  or
eqs.(\ref{eq:md1})-(\ref{eq:md4}) are represented by the $N\times N$
matrices
\begin{eqnarray}
 X^9(\sigma^1,\sigma^2)&\to& -2\pi iw_1L_1\p_{\theta^1}I_N
	+Y^1(\theta^1,\theta^2)\,,\label{eq:m9}\\
 X^8(\sigma^1,\sigma^2)&\to& -2\pi iw_2L_2\p_{\theta^2}I_N
	+\frac{w_2L_2}{N}\theta^1 I_N +Y^2(\theta^1,\theta^2),
	\label{eq:m8}\\
 X^m(\sigma^1,\sigma^2)&\to&X^m(\theta^1,\theta^2),\\
 A(\sigma^1,\sigma^2)&\to&A(\theta^1,\theta^2),
\end{eqnarray}
where ($\Xi$ represents $Y^1$, $Y^2$, $X^m$ and $A$)
\begin{equation}
  \Xi(\theta^1,\theta^2)=\sum_{u_1,u_2 \in \Z}
	\sum_{v_1,v_2=-M}^{M} \Xi_{(u_1N+v_1,u_2N+v_2)}\,
	e^{i(u_1N+v_1)\theta_1/N}\,e^{-i(u_2N+v_2)\theta_2/N}\,
	\lambda^{-v_1v_2/2}\,V^{v_2}\,U^{v_1}\,. \label{eq:Xi}
\end{equation}
Note that $\Xi(\theta^1,\theta^2)$ satisfies the boundary condition
\cite{UY4},
\begin{equation}
 \Xi(\theta^1+2\pi,\theta^2)=
	V\,\Xi(\theta^1,\theta^2)\,V^{\dagger},\quad
 \Xi(\theta^1,\theta^2+2\pi)=U\,\Xi(\theta^1,\theta^2)\,U^{\dagger}.
\end{equation}
Furthermore, the Poisson bracket and the double integrals are
represented as follows,
\begin{eqnarray}
 \acom{\cdot\,,\,\cdot} &\to&
	-i\frac{N}{2\pi}\,\com{\cdot\,,\,\cdot}\,,\label{eq:P2C} \\
 \int_0^{2\pi}d\sigma^1 d\sigma^2\,* &\to& \frac{1}{N}\int_0^{2\pi}
	d\theta^1d\theta^2\,\tr [\,*\,] \,. \label{eq:I2T}
\end{eqnarray}
Thus, from these results with a rescaling $\tau\to\tau/N$, the
matrix-regularized action of the wrapped membrane on $\R^9\times T^2$
is given by
\begin{eqnarray}
 S_{2+1}&=&\frac{LT}{2}\int d\tau\int_{0}^{2\pi}d\theta^1 d\theta^2
	\,\tr\Biggl[({F}_{\tau\theta^1})^2+({F}_{\tau\theta^2})^2
	-({F}_{\theta^1\theta^2 })^2\nn
 &&\qquad+ ({D}_\tau X^m)^2 -({D}_{\theta^1} X^m )^2
    -({D}_{\theta^2} X^m)^2+\frac{1}{2(2\pi L)^2}\,
	\com{X^m, X^n}^2\,\Biggr]\,,\quad\label{eq:S2+1}
\end{eqnarray}
where
\begin{eqnarray}
 F_{\tau\theta^1}&=&\p_{\tau} Y^1-\frac{w_1L_1}{L}\p_{\theta^1} A
	+\frac{i}{2\pi L}\com{A, Y^1}\,,\\
 F_{\tau\theta^2}&=&\p_{\tau} Y^2-\frac{w_2L_2}{L}\p_{\theta^2} A
	+\frac{i}{2\pi L}\com{A, Y^2}\,,\\
 F_{\theta^1\theta^2}&=&\frac{w_1w_2L_1 L_2}{NL}\,I_N
	+\frac{w_1L_1}{L}\,\p_{\theta^1}Y^2
	-\frac{w_2L_2}{L}\,\p_{\theta^2}Y^1
	+\frac{i}{2\pi L}\,\com{Y^1, Y^2}\,,\\
 D_{\tau}X^m&=&\p_{\tau} X^m +\frac{i}{2\pi L}\,\com{A,X^m}\,,\\
 D_{\theta^1} X^m&=&\frac{w_1L_1}{L}\,\p_{\theta^1} X^m
	+\frac{i}{2\pi L}\,\com{Y^1,X^m}\,,\\
 D_{\theta^2} X^m&=&\frac{w_2L_2}{L}\,\p_{\theta^2} X^m
	+\frac{i}{2\pi L}\,\com{Y^2,X^m}\,.
\end{eqnarray}
Note that the fields $Y^1, Y^2, X^m$ and $A$ have mass dimension $-1$
and the parameters $\tau, \theta^1, \theta^2$ have mass dimension $0$.
We also rewrite the action (\ref{eq:S2+1}) to the standard form of
Yang-Mills theory. In order to adjust the mass dimensions of the
fields and the parameters, we rewrite them by introducing some
dimensionful constants,
\begin{eqnarray}
 Y^1(\theta^1,\theta^2)&\to&\alpha A_1(x^1,x^2)\,,\label{eq:red1}\\
 Y^2(\theta^1,\theta^2)&\to&\alpha A_2(x^1,x^2)\,,\\
 X^m(\theta^1,\theta^2)&\to&\alpha \phi^m(x^1,x^2)\,,\\
 A(\theta^1,\theta^2)&\to&\alpha A_0(x^1,x^2)\,,\\
 \theta^1 &\to& x^1/\Sigma_1\,,\\
 \theta^2 &\to& x^2/\Sigma_2\,,\\
 \tau &\to& x^0/\Sigma\,,  \label{eq:red7}
\end{eqnarray}
where $\alpha$ has mass dimension $-2$ and $\Sigma_1, \Sigma_2$ and
$\Sigma$ have mass dimension $-1$.
Then, the action (\ref{eq:S2+1}) is rewritten by
\begin{eqnarray}
S_{2+1}&=&\frac{LT}{2}\frac{1}{\Sigma_1\Sigma_2\Sigma}
    \int\!dx^0\!\!\int_0^{2\pi\Sigma_1}\!\!\!dx^1\!\!
	\int_0^{2\pi\Sigma_2}\!\!\!dx^2\ \mbox{Tr}
	\Biggr[ (F_{\tau\theta^1})^2+(F_{\tau\theta^2})^2
	-(F_{\theta^1\theta^2})^2\nn
 &&\qquad{}+(D_{\tau}X^m)^2-(D_{\theta^1}X^m)^2 -(D_{\theta^2}X^m)^2
    +\frac{\alpha^4}{2(2\pi L)^2}\,\com{\phi^{m},\phi^{n}}^2\Biggr],\\
 &&F_{\tau\theta^1}=\Sigma\alpha\p_0 A_1
    -\frac{L_1}{L}\,\Sigma_1\alpha\p_{1}A_0
	+i\frac{\alpha^2}{2\pi L}\,\com{A_0,A_1},\label{Fie1}\\
 &&F_{\tau\theta^2}=\Sigma\alpha\p_0 A_2
    -\frac{L_2}{L}\,\Sigma_2\alpha\p_{2}A_0
	+i\frac{\alpha^2}{2\pi L}\,\com{A_0,A_2},\label{Fie2}\\
 &&F_{\theta^1\theta^2}=\frac{L_1L_2}{NL}I_N
    +\frac{L_1}{L}\,\Sigma_1\alpha\p_{1} A_2
	-\frac{L_2}{L}\,\Sigma_2\alpha\p_{2}A_1
	+i\frac{\alpha^2}{2\pi L}\,\com{A_1,A_2},\label{Fie3}\\
 &&D_{\tau}X^m=\Sigma\alpha\p_0\phi^m
	+i\frac{\alpha^2}{2\pi L}\,\com{A_0,\phi^m},\\
 &&D_{\theta^1}X^m=\frac{L_1}{L}\,\Sigma_1\alpha\p_{1}\phi^m
	+i\frac{\alpha^2}{2\pi L}\,\com{A_1,\phi^m},\\
 &&D_{\theta^2}X^m=\frac{L_2}{L}\Sigma_2\alpha\p_{2} \phi^m
	+i\frac{\alpha^2}{2\pi L}\,\com{A_2,\phi^m},\label{DX3}
\end{eqnarray}
where $\p_0\equiv \p/\p x^0$ and $\p_i\equiv \p/\p x^i$.
Here we have put $w_1=w_2=1$ for simplicity.
In order to bring the field strength (\ref{Fie1})-(\ref{Fie3})
into the standard form, we obtain the following relations \cite{UY4}
\footnote{Eqs.(\ref{eq:Td1}) and (\ref{eq:Td2}) represent the
T-duality which relates the radii ($L_1,L_2$) of the 2-torus in
M-theory and the those ($\Sigma_1,\Sigma_2$) of the dual 2-torus in
the super Yang-Mills theory.
We should stress that we have obtained the same relations
from the different viewpoint \cite{UY4}.}
\begin{eqnarray}
  \Sigma&=&\frac{\alpha}{2\pi L}\,,\label{eq:Td0}\\
  \Sigma_1&=&\frac{\alpha}{2\pi L_1}\,,\label{eq:Td1}\\
  \Sigma_2&=&\frac{\alpha}{2\pi L_2}\,.\label{eq:Td2}
\end{eqnarray}
Then, we have obtained the standard form of a bosonic part of
(2+1)-dimensional maximally supersymmetric $U(N)$ Yang-Mills theory
with the constant magnetic flux,
\begin{eqnarray}
 S_{2+1}&=&\frac{1}{2\gym^2}\int dx^0\!
    \int_0^{2\pi\Sigma_1}\!\!dx^1\!
    \int_0^{2\pi\Sigma_2}\!\!dx^2\,\mbox{Tr}\Biggl[
	(F_{01})^2+(F_{02})^2 -(F_{12})^2\nn
 &&\qquad{}+(D_{0}\phi^m)^2-(D_{1}\phi^m)^2 -(D_{2}\phi^m)^2
	+\frac{1}{2}\com{\phi^m,\phi^n}^2\Biggr], \label{eq:SYM}\\
 F_{01}&=&\p_0 A_1 -\p_{1}A_0 +i\com{A_0,A_1},\label{Fie1'}\\
 F_{02}&=&\p_0 A_2 -\p_{2}A_0+i\com{A_0,A_2},\label{Fie2'}\\
 F_{12}&=&f_{12}+\p_{1} A_2 -\p_{2}A_1+i\com{A_1,A_2},\label{Fie3'}\\
 D_{\alpha}\phi^m&=&\p_\alpha \phi^m
	+i\com{A_\alpha,\phi^m},\label{DX3'}\qquad (\alpha=0,1,2)
\end{eqnarray}
with the boundary conditions ($\Xi$ stands for $A_\alpha$
and $\phi^m$),
\begin{eqnarray}
 \Xi(x^1+2\pi\Sigma_1,x^2)&=&V\Xi(x^1,x^2)V^{\dagger}\,,
	\label{eq:BC1a}\\
 \Xi(x^1,x^2+2\pi\Sigma_2)&=&U\Xi(x^1,x^2)U^{\dagger}\,,
	\label{eq:BC1b}
\end{eqnarray}
where the constant magnetic flux $f_{12}$ is given by
\begin{equation}
	f_{12}=\frac{1}{2\pi N\Sigma_1\Sigma_2}\,I_N\,,
\end{equation}
and $\gym$ is the gauge coupling constant of mass dimension one
half, which is given by
\begin{equation}
  \gym^2=(2\pi)^{-2}(\Sigma_1\Sigma_2)^{-1/2}\,(L_1L_2)^{-3/2}
	\,T^{-1}.\label{eq:GYM}
\end{equation}
We also define the dimensionless gauge coupling constant $\ogym$ by
\begin{eqnarray}
 \ogym^2\equiv \gym^2(2\pi\Sigma_12\pi\Sigma_2)^{1/2}
	=(2\pi)^{-1}(L_1L_2)^{-3/2}\,T^{-1}
	=\frac{2\pi l_{11}^3}{(L_1L_2)^{3/2}}\,, \label{eq:DLGYM}
\end{eqnarray}
where $l_{11}$ is the 11-dimensional Planck length related to $T$ by
$T^{-1}=(2\pi)^2 l_{11}^3$.
This dimensionless gauge coupling constant exactly agrees with that
obtained in ref.\cite{FHRS} including the numerical
constant.\footnote{Note that the parameters $\Sigma_1,\Sigma_2$ and
$L_1,L_2$ in ref.\cite{FHRS} represent the circumferences but not the
radii.}
Note that in refs.\cite{FHRS} the super Yang-Mills theory was regarded
as the low energy effective theory of D-branes in deriving such a
relation, while we have taken a different approach of matrix
regularization of supermembrane in this section.
Furthermore, the constant magnetic flux $f_{12}$ in eq.(\ref{Fie3'})
agrees with that obtained in refs.\cite{GRT,FHRS} including the
numerical constant.

\subsection{The general matrix representation}
We have adopted a simple representation
(\ref{eq:mx1})-(\ref{eq:mx3}) for the star-commutator algebra
(\ref{eq:al1})--(\ref{eq:al4}) to bring the wrapped supermembrane
action to the standard form of super Yang-Mills theory in
eqs.(\ref{eq:SYM})-(\ref{DX3'}).
However, we could adopt more general representation of the algebra
\begin{eqnarray}
 e^{ik_1 \sigma^1+ik_2 \sigma^2} &\to&
   e^{ik_i T^i_{~j}\,\theta^j/N }
	\lambda^{-v_1v_2/2}\,V^{v_2}\,U^{v_1}, \label{eq:gr1}\\
 \sigma^2 &\to& c^i \p_{\theta^i} I_N
	+d_i\theta^i I_N,\label{eq:gr2} \\
 \sigma^1 &\to& e^i \p_{\theta^i} I_N
	+f_i \theta^i I_N\,. \label{eq:gr3}
\end{eqnarray}
In fact, we can easily check that this is also a representation
of the star-commutator algebra with following constraints
\begin{eqnarray}
  ik_i\,T^i_{~j}\,e^j &=& -2\pi k_2\,, \\
  ik_i\,T^i_{~j}\,c^j &=& 2\pi k_1\,, \\
  e^i d_i -c^i f_i &=& \frac{2\pi i}{N}\,,
\end{eqnarray}
where the matrix $T^i_{~j}$ is given by
\begin{eqnarray}
 T^i_{~j}&=& \frac{2\pi i}{(c^1e^2-c^2e^1)}
    \left( \begin{array}{@{\,}cc@{\,}}
      -e^2  &  e^1 \\ -c^2  &  c^1 \end{array} \right)\,.
\end{eqnarray}
Note that in such general representation the resultant action
is not always in the standard form of the super Yang-Mills action.
In section \ref{S:YMPQ} we shall consider the supermembrane wrapped
around the general two-cycles of $T^2$. Then we shall use this general
representation to bring the wrapped supermembrane action into the
standard super Yang-Mills action.

\section{Wrapped supermembrane in curved
	background}\label{S:L}
In this section we consider the supermembrane wrapped around
nontrivial two cycles of $T^2$ and apply matrix regularization
procedure to it.
Then we perform the double dimensional reduction and derive
the matrix $(p,q)$-strings.

\subsection{Setup}
The bosonic part of the lightcone supermembrane in 11-dimensional
curved background is given in ref.\cite{DWPP2,DWPP3}.
It was conjectured in ref.\cite{CDS} to identify a
lightcone component of the background 3-form $A_{-MN}$ with the
noncommutative parameter of the 2-torus.
We need more study on this issue, however, since our goal in this
section is to deduce the Green-Schwarz $(p,q)$-string action from the
matrix-regularized wrapped supermembrane by the double dimensional
reduction, we shall put the background fields along the lightcone
directions zero.
Then the action in ref.\cite{DWPP2} is reduced to contain only fields
with the transverse indices,
\begin{eqnarray}
  S&=&\frac{LT}{2}\int\!\! d\tau\!
 	\int^{2\pi}_0\!\!d\sigma^1d\sigma^2\,\Bigg[(D_\tau X^M)^2\nn
   &&\qquad-\frac{1}{2L^2}\acom{X^M,X^N}^2+\frac{1}{L}
	D_\tau X^M A_{MNP}\acom{X^N,X^P} \Bigg]\,, \label{eq:MLCC}
\end{eqnarray}
where $A_{MNP}$ is the 3-form field, $X^M$ is target-space coordinates
and the transverse indices $M,N,P=1,2,\cdots,9$ are contracted by the
target-space metric $G_{MN}$.
Considering the line element on a 2-torus
\begin{equation}
 ds^2_{T^2}=G_{uv}\,dX^udX^v=\Big(G_{88}
    -\frac{(G_{89})^2}{G_{99}}\Big)(dX^8)^2+G_{99}
    \Big(dX^9+\frac{G_{89}}{G_{99}}dX^8\Big)^2\,,\label{eq:dst2}
\end{equation}
where $u,v=8,9$, we shall choose the target-space coordinates
satisfying the following boundary conditions \cite{OUY}
\begin{eqnarray}
 \sqrt{G_{99}}\,X^{9}(\sigma^1,\sigma^2+2\pi)&=&2\pi w_1 L_1 p
    +\sqrt{G_{99}}\,X^{9}(\sigma^1,\sigma^2)\,,\label{eq:pq1p}\\
	\sqrt{G_{88}-\frac{(G_{89})^2}{G_{99}}}\,
	X^{8}(\sigma^1,\sigma^2+2\pi)
  &=&2\pi w_1 L_2 q +\sqrt{G_{88}-\frac{(G_{89})^2}{G_{99}}}
	\,X^{8}(\sigma^1,\sigma^2)\,,\label{eq:pq1q}\\
  \sqrt{G_{99}}\,X^{9}(\sigma^1+2\pi,\sigma^2)&=& 2\pi w_2 L_1 r
    +\sqrt{G_{99}}\,X^{9}(\sigma^1,\sigma^2)\,,\label{eq:pq2r}\\
	\sqrt{G_{88}-\frac{(G_{89})^2}{G_{99}}}\,
	X^{8}(\sigma^1+2\pi,\sigma^2)
  &=& 2\pi w_2 L_2 s +\sqrt{G_{88}-\frac{(G_{89})^2}{G_{99}}}
	\,X^{8}(\sigma^1,\sigma^2)\,.\label{eq:pq2s}
\end{eqnarray}
or
\begin{eqnarray}
 X^9(\sigma^1,\sigma^2) &=& R_1\,(w_1p\sigma^2 + w_2r\sigma^1)
    + Y^1(\sigma^1,\sigma^2)\,,\label{eq:md1'}\\
 X^8(\sigma^1,\sigma^2)&=&R_2\,(w_1q\sigma^2+w_2s\sigma^1)
	+Y^2(\sigma^1,\sigma^2)\,,\label{eq:md2'}
\end{eqnarray}
where\footnote{We may assume $\Nis>0$ and $w_1>0$ without loss of
generality since we can flip the signs of $(p,q)\to(-p,-q)$ (for
$w_1$) and $(r,s)\to(-r,-s)$ (for $\Nis$) if necessary.
Furthermore, we may see that eq.(\ref{eq:pq}) leads to $(r,s)=n(-q,p)$
($n\in\N$).\label{f:1}}
\begin{equation}
 pr+qs=0,\quad ps-qr\equiv\Nis>0,\quad (p,q,r,s \in\Z,
	\ w_1\in\N\backslash\{0\}\,,
	\ w_2\in\Z\backslash\{0\})\label{eq:pq}
\end{equation}
and\footnote{We shall see $R_1=L_1\,e^{-2\phi/3}$ from
eq.(\ref{eq:KKmetric}) and hence M/IIA-relation, or
11d/IIA-SUGRA-relation, leads to $R_1=\ell_{11}$ (11-dimensional
Planck length).\label{fn:1}}
\begin{equation}
 R_1\equiv \frac{L_1}{\sqrt{G_{99}}}\,,\quad R_2\equiv
	\frac{L_2}{\sqrt{G_{88}-\frac{(G_{89})^2}{G_{99}}}}\,.
\end{equation}
$Y^i\ (i=1,2)$ and the other fields in eqs.(\ref{eq:md3}) and
(\ref{eq:md4}) satisfy the periodic boundary conditions,
$Y^1(\sigma^1+2\pi,\sigma^2)=Y^1(\sigma^1,\sigma^2+2\pi)
=Y^1(\sigma^1,\sigma^2)$, etc..
The above expressions represent that the supermembrane is wrapping
$w_1p$-times around one of the two compact directions ($X^9$) and
$w_1q$-times around the other direction ($X^8$), or $w_1$-times
around $(p,q)$-cycle along the $\sigma^2$-direction on the worldsheet.
And also it is wrapping $w_2$-times around $(r,s)$-cycle along the
$\sigma^1$-direction.
These two cycles are orthogonal to each other and intersect at least
once. Thus, this wrapped supermembrane is expected to give the
$(p,q)$-string \cite{Sch,OUY}.
In fact, we shall see below that the $(p,q)$-string comes
out through the double dimensional reduction.

\subsection{Matrix regularization and double dimensional reduction}
We shall follow the matrix regularization procedure presented in
section \ref{S:MS} \cite{UY4} and then consider the double dimensional
reduction \cite{DHIS} with the matrices.
One comment is in order: In this paper we do not consider the
matrix regularization of the background fields, which play
the role of probing the membrane $X^M$ and hence the background
fields $G_{MN}$ and $A_{MNP}$ are proportional to the unit matrix in
the matrix regularized action.
The double dimensional reduction is carried out along the
$(p,q)$-cycle \cite{OUY}, however, we should be careful to really
deduce the ($p,q$)-string.
First we should notice the followings. Once we intend to deduce type
IIB superstring we shall consider the shrinking volume limit of the
2-torus keeping the ratio of the radii finite,
\begin{equation}
  \frac{L_1}{L_2}\equiv \gIIB : \mathrm{finite}.\quad (L_1,L_2\to0)
\end{equation}
and the ratio is the type IIB coupling constant \cite{Asp}.
On the other hand, by using the relations between the 11-dimensional
supergravity and 9-dimensional type IIB background fields in
eqs.(\ref{eq:2BM1})-(\ref{eq:2BM3}) \cite{BHO,MO} we have
\begin{equation}
  \sqrt{\frac{G_{88}-\frac{(G_{89})^2}{G_{99}}}{G_{99}}}
  =  e^{-\varphi}=\frac{1}{\gIIB}=\frac{L_2}{L_1}\,,\label{eq:gbLL}
\end{equation}
where $\varphi$ is a background of the type IIB dilaton.
Thus eq.(\ref{eq:gbLL}) leads to
\begin{equation}
  R_1 = R_2 \equiv R_B\,.\label{eq:RB}
\end{equation}
Then we set (\ref{eq:RB}) hereafter in this section.\footnote{We do
not set eq.(\ref{eq:RB}) in section \ref{S:YMPQ}.}

Next we determine the spacetime directions to align with the
worldvolume coordinate, or we fix the gauge.
We define $X^y$ and $X^z$ by an $SO(2)$ rotation of the target-space
\cite{OUY}
\begin{equation}
  \left(\begin{array}{@{\,}c@{\,}} X^z\\ X^y\end{array} \right)=
   \Opq \left(\begin{array}{@{\,}c@{\,}}X^9\\ X^8\end{array}\right),
	\label{eq:Opq}
\end{equation}
where
\begin{equation}
  \Opq =\frac{1}{\cpq}\left(\begin{array}{@{\,}cc@{\,}}
	p & q \\[5pt] -q & p \end{array}\right)
  \equiv\left(\begin{array}{@{\,}cc@{\,}}
        \hat{p} & \hat{q} \\[5pt]
	-\hat{q} & \hat{p} \end{array}\right)\in SO(2),
 \quad\cpq\equiv \sqrt{p^2 +q^2}\,.\label{eq:so2mx}
\end{equation}
Then $X^z$- and $X^y$-directions are given by
\begin{eqnarray}
 X^z &=&w_1\cpq R_B \sigma^2+\hp\,Y^1+\hq\,Y^2
	\equiv C_1\,\sigma^2+Y^z\,,\label{eq:X^z} \\
 X^y &=& \frac{w_2\Nis R_B}{\cpq}\sigma^1-\hq\,Y^1+\hp\,Y^2
	\equiv C_2\,\sigma^1+Y^y\,,\label{eq:X^y}
\end{eqnarray}
and they are aligned with ($p,q$)- and ($r,s$)-cycles, respectively.
The transverse metric and 3-form are transformed as
\begin{equation}
 \tG_{UV} = G_{MN}\,\frac{\p X^M}{\p X^U}\,
	\frac{\p X^N}{\p X^V}\,,\quad
   \tA_{UVW}= A_{MNP}\,\frac{\p X^M}{\p X^U}\,
    \frac{\p X^N}{\p X^V}\,\frac{\p X^P}{\p X^W}\,,
\end{equation}
where $U,V,W=1,2,\cdots,7,y,z$. We shall parametrize $\tG_{UV}$ as
(cf. eq.(\ref{eq:KKmetric}))
\begin{equation}
 \tG_{UV} =\left(\begin{array}{@{\,}ccc@{\,}}
	\frac{1}{\sqrt{\tG_{zz}}}\,\tg_{mn}
	+\frac{1}{\tG_{zz}}\tG_{mz}\tG_{nz}&
	\frac{1}{\sqrt{\tG_{zz}}}\,\tg_{m y}
	+\frac{1}{\tG_{zz}}\tG_{mz}\tG_{yz}& \tG_{mz} \\[10pt]
	\frac{1}{\sqrt{\tG_{zz}}}\,\tg_{yn}
	+\frac{1}{\tG_{zz}}\tG_{yz}\tG_{\nu z}&
	\frac{1}{\sqrt{\tG_{zz}}}\,\tg_{yy}
	+\frac{1}{\tG_{zz}}\tG_{yz}\tG_{yz}& \tG_{yz} \\[10pt]
	\tG_{\nu z}&\tG_{yz} & \tG_{zz}	\end{array}\right).
\end{equation}

Here we shall introduce a noncommutativity on spacesheet
(\ref{eq:al1})-(\ref{eq:al4}) and give a matrix representation as is
given in eqs.(\ref{eq:mx1})-(\ref{eq:mx3}), (\ref{eq:P2C}) and
(\ref{eq:I2T}).
The double dimensional reduction on the matrices is carried out by
imposing the following conditions on the oscillators ($\Xi$ stands for
$X^m, Y^y$ and $A$) and background fields,
\begin{equation}
	Y^z=0,\quad \p_{\theta^2} \Xi=0\,,\label{eq:DDR1}
\end{equation}
and
\begin{equation}
  \p_{\theta^2} \tG_{UV}=\p_{\theta^2} \tA_{UVW}=0\,.\label{eq:DDR2}
\end{equation}
Then the oscillators are reduced to the diagonal matrices
\begin{equation}
  \Xi(\theta^1)=\sum_{u_1 \in \Z}
	\sum_{v_1=-M}^{M} \Xi_{(u_1N+v_1,0)}\,
	e^{i(u_1N+v_1)\theta^1/N}\,U^{v_1}\,,\label{eq:XiDDR}
\end{equation}
and the commutators between them automatically vanish.
Under the double dimensional reduction, the non-zero components in the
action come from the followings,
\begin{eqnarray}
 D_\tau X^z&=&-\frac{C_1}{L}\,\p_{\theta^1} A\,, \\
 D_\tau X^y&=&\p_\tau Y\,, \\
 D_\tau X^m&=&\p_\tau X^m\,, \\
 \frac{-i}{2\pi L} \com{X^z,X^y}&=&-\frac{C_1}{L}\,\p_{\theta^1}Y\,,\\
 \frac{-i}{2\pi L} \com{X^z,X^m}&=&-\frac{C_1}{L}\,\p_{\theta^1}X^m\,,
\end{eqnarray}
where $Y \equiv Y^y+(C_2\,\theta^1/N)I_N$.
Then the action (\ref{eq:MLCC}) is rewritten by
\begin{eqnarray}
 S&=&\frac{2\pi TL}{2}\int d\tau\!\int^{2\pi}_0\!d\theta^1~\tr\Bigg[
    \Big(\frac{\tilde{g}_{mn}}{\sqrt{\tilde{G}_{zz}}}
    +\frac{\tG_{mz}\tG_{nz}}{\tilde{G}_{zz}}\Big)
	\p_0{X}^m\p_0{X}^n \nn
 &&{}+2\Big(\frac{\tg_{ym}}{\sqrt{\tG_{zz}}}
    +\frac{\tG_{mz}\tG_{yz}}{\tilde{G}_{zz}}\Big)
	\p_0{X}^m\p_0Y\nn
 &&{}-\frac{2C_1}{L}\Big(\tG_{mz}\p_0{X}^m\p_1A
	+\tG_{yz}\p_0Y\p_1A \Big)
	+\tG_{yy}(\p_0Y)^2
    +\left(\frac{C_1}{L}\right)^2\tG_{zz}(\p_1 A)^2\nn
 &&{}-\left(\frac{C_1}{L}\right)^2\sqrt{\tG_{zz}}
    \Big(\tilde{g}_{yy}(\p_1Y)^2
    +2\tilde{g}_{my}\p_1 X^m\p_1Y
	+\tilde{g}_{mn}\p_1X^m\p_1X^n\Big)\nn
 &&{}+\frac{2C_1}{L}(A_{nmz}\p_0{X}^n\p_1 X^m
	+A_{mzy}\epn^{ab}\p_a Y\p_b X^m) \Bigg],
\end{eqnarray}
where $a,b=0,1$ and we adopt the notation of $(\p_0,\p_1)\equiv
(\ptau,\p_{\theta^1})$ only in section \ref{S:L}.
Then, solving the field equation of $A$ and rescaling
$\tau\to\tau L/(C_1\sqrt{\tG_{zz}})$
we obtain the double dimensionally reduced action
\begin{eqnarray}
 S&=&\frac{2\pi T}{2}\int d\tau \int^{2\pi}_0 d\theta^1
	C_1 \tr \Bigg[\eta^{ab}( \tilde{g}_{mn}\p_a{X}^m\p_b{X}^n
	+2\tg_{my}\p_a{X}^m\p_b{Y}+\tg_{yy}\p_a{Y}\p_b{Y})\nn
 &&{}+2(\tA_{nmz}\ptau{X}^n\p_1 X^m
    +\tA_{mzy}\epn^{ab}\p_a Y \p_b X^m)\Bigg]\,.\label{eq:Macr}
\end{eqnarray}

\subsection{$(p,q)$-string from wrapped supermembrane}
In this subsection we derive the $(p,q)$-string action from the
reduced supermembrane action in eq.(\ref{eq:Macr}).
The action has an abelian isometry associated with the other
compactified $Y$-direction, we can make a dual transformation as is
the case with sigma models.
Introducing a variable $\tilde{Y}$, which is seen to be dual to $Y$,
eq.(\ref{eq:Macr}) can be rewritten in a classically equivalent form
\begin{eqnarray}
 S&=&\frac{2\pi T}{2}\int d\tau\!\int^{2\pi}_0\!d\theta^1\,C_1\,
  \tr\Bigg[\eta^{ab}(\tilde{g}_{mn}\p_a{X}^m\p_b{X}^n
    +2\tg_{my}\p_a{X}^m G_b+\tg_{yy}G_a G_b)\nn
 &&+2(\tA_{nmz}\ptau{X}^n\p_1 X^m +\tA_{mzy}\epn^{ab}G_a \p_b X^m)
	+2 \epn^{ab}\tilde{Y} \p_a G_b \Bigg]\,,\label{eq:d-Macr}
\end{eqnarray}
since the variation w.r.t.\ $\tilde{Y}$ leads to $\epn^{ab}\p_aG_b=0$
or $G_a=\p_a Y$ and hence eq.(\ref{eq:Macr}) can be
reproduced.\footnote{We assume that the background fields are
independent of $\tilde{Y}$ in eq.(\ref{eq:d-Macr}).}
On the other hand, assuming that all the fields are independent of
$G_a$ (or $Y$), the variation w.r.t.\ $G_a$ leads to
\begin{eqnarray}
 G_a&=&\frac{1}{\tg_{yy}}
    \Bigl\{\eta_{ab}\epn^{cb}\p_c \tilde{Y}-\tg_{my}\p_aX^m-A_{mzy}
    (\p_1 X^m \eta_{0a}-\p_0{X}^m\eta_{a1})\Bigr\}\,,
\end{eqnarray}
and hence we have
\begin{eqnarray}
 S&=&\frac{2\pi T}{2}\int d\tau\!\int^{2\pi}_0\!d\theta^1\,C_1\,
    \tr\Bigg[\Big(\tg_{mn} -\frac{\tg_{my}\tg_{ny}
    -\tA_{mzy}\tA_{nzy}}{\tg_{yy}}\Big)\eta^{ab}\p_a X^m \p_b X^n\nn
 &&{}+2\,\frac{\tA_{mzy}}{\tg_{yy}}\,\eta^{ab}\p_aX^m \p_b\tilde{Y}
	+\frac{1}{\tg_{yy}}\,\eta^{ab}\p_a \tilde{Y} \p_b\tilde{Y}\nn
 &&{}+\Big(\tA_{mnz}+2\,\frac{\tA_{mzy} \tg_{ny}}{\tg_{yy}}\Big)
	\epn^{ab}\p_a X^m \p_b X^n+2\,\frac{\tg_{my}}{\tg_{yy}}
	\epn^{ab}\p_a \tilde{Y} \p_b X^m\Bigg]\,.\label{eq:ddrac}
\end{eqnarray}

Now that we consider T-dual for the background fields in
eq.(\ref{eq:Macr}) (or eq.(\ref{eq:ddrac})).
Since we regard $X^{9}$ ({\sl not} $X^{z}$) as the 11th direction,
we should take T-dual along the $X^8$-direction to transform type IIA
superstring theory to type IIB superstring theory.
Then we can rewrite the background fields in terms of those of the
type IIB supergravity as follows \cite{BHO,MO}(cf. Appendix \ref{S:R}),
\begin{eqnarray}
 \tg_{m y}&=&\frac{B^{(pq)}_{8m}}{\jmath_{88}}\,, \\
 \tg_{yy}&=&\frac{1}{\jmath_{88}\sqrt{(\hp+\hq l)^2
		+ e^{-2\varphi}\hq^2}}\,,\\
 \tg_{mn}&=&\sqrt{(\hp+\hq l)^2
	+e^{-2\varphi}\hq^2}\,\left(\jmath_{mn}
	-\frac{\jmath_{8m}\jmath_{8n}}{\jmath_{88}}
    +\frac{B^{(pq)}_{8m}B^{(pq)}_{8n}}{\jmath_{88}}\right) \,,\\
 \tA_{mnz}&=&\sqrt{(\hp+\hq l)^2+e^{-2\varphi}\hq^2}\,
    \Bigl(B^{(pq)}_{mn}
    +\frac{2}{\jmath_{88}}B^{(pq)}_{8[m}\jmath_{n]8}\Bigr)\,,\\
 \tA_{myz}&=&-\frac{\jmath_{8m}}{\jmath_{88}}=\sqrt{(\hp+\hq l)^2
	+e^{-2\varphi}\hq^2}\,\tilde{g}_{yy} \jmath_{8m}\,,
\end{eqnarray}
where $\BNS_{IJ}$ and $\BR_{IJ}$ are the NSNS and RR
second-rank antisymmetric tensors, respectively, $\jmath_{IJ}$
are the metric in type IIB supergravity, $l=G_{89}/G_{99}=A_8$
and
\begin{equation}
 B^{(pq)}_{IJ}=\frac{\hp\,\BNS_{IJ}+\hq\,\BR_{IJ}}{%
	\sqrt{(\hp+\hq l)^2+e^{-2\varphi}\hq^2}}\,,
\end{equation}
where $I,J=1,2,\cdots,8$.
Then, plugging these equations into eq.(\ref{eq:ddrac}) we have
\begin{eqnarray}
 S&=&\frac{2\pi T}{2}\int d\tau\!\int^{2\pi}_0\!d\theta^1
    \,C_1\sqrt{(\hp+\hq l)^2+e^{-2\varphi}\hq^2}\nn
  &&{}\qquad\times\tr\Bigl[\eta^{ab}\p_a\tilde{X}^I\p_b \tilde{X}^J
    \jmath_{IJ}+\epn^{ab}\p_a \tilde{X}^I \p_b
    \tilde{X}^JB^{(pq)}_{IJ}\Bigr]\,,\label{eq:pqac}
\end{eqnarray}
where we have defined $\tilde{X}^I\equiv (X^m,\tilde{Y})$.
Once we regard $X^{9}$ as the 11th direction, the type IIA string
tension $T_s$ is given by $2\pi L_1T/\sqrt{G_{99}}$ \cite{Sch}
since the 11-dimensional metric $G_{MN}$ is converted to the type
IIA metric $g_{IJ}$ by the relation
$G_{IJ}=g_{IJ}/\sqrt{G_{99}}$.
Also, if we assume that $l$ and $\varphi$ are constant and hence
$e^\varphi=\gIIB$, we have
\begin{equation}
 2\pi TC_1\sqrt{(\hp+\hq l)^2+e^{-2\varphi}\hq^2}
	=w_1T_s\sqrt{(p+ql)^2+e^{-2\varphi}q^2}\equiv w_1T_{pq}
\end{equation}
where $T_{pq}$ is the tension of a $(p,q)$-string in type IIB
superstring theory \cite{Sch}.
In particular, we see that both of the NSNS and RR antisymmetric
tensors couple to $\tilde{X}^I$ in eq.(\ref{eq:pqac}),
which implies that the reduced action (\ref{eq:pqac}) is, in fact,
that of the $(p,q)$-strings.
Note that $w_1$ is just the number of copies of the resulting
$(p,q)$-string. If we set $q$ to be zero and hence take
$(p,q,r,s)=(1,0,0,1)$, we have the fundamental strings in type IIB
superstring theory. On the other hand, $(p,q,r,s)=(0,1,1,0)$ leads to
the strings which couple minimally with the RR B-field, i.e., the
D-strings.

\section{Matrix-regularized action in curved
	background}\label{S:YMPQ}
In this section we perform the matrix regularization on the wrapped
supermembrane in curved background by adopting a suitable choice of
matrix representation. As we mentioned before, the background fields
are to be proportional to the unit matrix in the matrix regularization
here.
Then we derive the standard form of (2+1)-dimensional super Yang-Mills
action in the curved background.

\subsection{Standard form of super Yang-Mills action}
Let us start with the wrapped supermembrane (\ref{eq:MLCC}) where
$X^9,X^8,X^m$ and $A$ are given by eqs.(\ref{eq:md1'}),
(\ref{eq:md2'}), (\ref{eq:md3}) and (\ref{eq:md4}),
respectively.
In this section we assume that the background metric is block-diagonal
\begin{equation}
  G_{MN}=\left(\begin{array}{@{\,}cc@{\,}}
    G_{mn} & 0 \\ 0 & G_{uv}\\ \end{array}\right)\,.\label{eq:blockG}
\end{equation}
We shall perform the matrix regularization.
The procedure is the same as before (cf. subsection \ref{S:MR}),
however, we should be careful in the choice of matrix
representation of the algebra.
In this case by using the general matrix representation in
eqs.(\ref{eq:gr1})-(\ref{eq:gr3}), we shall search for the set of
parameters which makes the matrix representations of $X^9$ and $X^8$
similar to eqs.(\ref{eq:m9}) and (\ref{eq:m8}), respectively.
In fact, we find that the choice of the parameters
\begin{eqnarray}
 &&c^1=-\frac{2\pi i s}{w_1\Nis},\quad c^2=\frac{2\pi i r}{w_1\Nis}\,,
    \quad e^1=\frac{2\pi iq}{w_2\Nis}\,,
    \quad e^2=-\frac{2\pi i p}{w_2\Nis}\,, \nn
 &&d_1=-\frac{w_2r}{N}\,,\quad d_2=0,\quad f_1=\frac{w_1p}{N}\,,
    \quad f_2=0,\label{eq:GMRSC}
\end{eqnarray}
leads to the matrix representation
\begin{eqnarray}
 X^9(\sigma^1,\sigma^2)&\to& -2\pi iR_1\p_{\theta^1}I_N
	+Y^1(\theta^1,\theta^2)\,,\label{eq:hm9}\\
 X^8(\sigma^1,\sigma^2)&\to& -2\pi iR_2\p_{\theta^2}I_N
	+\frac{w_1w_2\Nis R_2}{N}\,\theta^1 I_N
	+Y^2(\theta^1,\theta^2)\,,\label{eq:hm8}\\
 X^m(\sigma^1,\sigma^2)&\to&X^m(\theta^1,\theta^2)\,,\\
 A(\sigma^1,\sigma^2)&\to&A(\theta^1,\theta^2)\,,
\end{eqnarray}
and the oscillation modes of the $N\times N$ matrices
are give by ($\Xi\,(\theta^1,\theta^2)$ stands for
$Y^1\,(\theta^1,\theta^2)$, $Y^2\,(\theta^1,\theta^2)$,
$X^m\,(\theta^1,\theta^2)$ and $A\,(\theta^1,\theta^2)$)
\begin{eqnarray}
  \Xi(\theta^1,\theta^2)&=&\sum_{u_1,u_2\in\Z}
	\sum_{v_1,v_2=-M}^{M}\!\!\Xi_{(u_1N+v_1,u_2N+v_2)}\,
  e^{iK_1\theta^1/N} e^{-iK_2\theta^2/N}\,
    \lambda^{-v_1v_2/2}\,V^{v_2}U^{v_1},\quad\label{eq:Xipq}
\end{eqnarray}
where
\begin{equation}
  K_1=w_1p(u_1N+v_1)-w_2r(u_2N+v_2)\,,\quad
	K_2=-w_1q(u_1N+v_1)+w_2s(u_2N+v_2)\,.\label{eq:Ks}
\end{equation}
Then, $\Xi$ satisfies the boundary conditions
\begin{eqnarray}
 \Xi(\theta^1+\frac{2\pi s}{w_1\Nis},
	\,\theta^2-\frac{2\pi r}{w_1\Nis})
  &=&V\,\Xi(\theta^1,\theta^2)\,V^\dagger\,,\label{eq:pqBC1}\\
 \Xi(\theta^1-\frac{2\pi q}{w_2\Nis},
	\,\theta^2+\frac{2\pi p}{w_2\Nis})
  &=&U\,\Xi(\theta^1,\theta^2)\,U^\dagger\,,\label{eq:pqBC2}
\end{eqnarray}
or
\begin{eqnarray}
 \Xi(\theta^1+2\pi,\theta^2)&=&
      U^{w_2r}V^{w_1p}\,\Xi(\theta^1,\theta^2)
	\,(U^{w_2r}V^{w_1p})^\dagger\,,\label{eq:pqBC1n}\\
 \Xi(\theta^1,\theta^2+2\pi)&=&U^{w_2s}V^{w_1q}\,
    \Xi(\theta^1,\theta^2)\,(U^{w_2s}V^{w_1q})^\dagger\,,
      \label{eq:pqBC2n}
\end{eqnarray}
which express that the supermembrane wraps around
($p,q$)- and ($r,s$)-cycles.
Furthermore, since we have
\begin{eqnarray}
 &&\scom{\sigma^1,\sigma^2}\to
    \com{c^i \p_{\theta^i}+d_i\theta^i,e^i\p_{\theta^i}+f_i\theta^i}
	=\frac{2\pi i}{N}\,,\\
 &&\frac{1}{N} \int_F d\theta^1d\theta^2\,\tr\,I_N
	=\frac{(2\pi)^2}{w_1|w_2|\Nis}\,,
\end{eqnarray}
where $F$ is a parallelogram generated by the
two vectors, $(2\pi s/(w_1\Nis), -2\pi r/(w_1\Nis))$ and
$(-2\pi q/(w_2\Nis), 2\pi p/(w_2\Nis))$,
the Poisson bracket and the double integral are represented as
\begin{eqnarray}
 \acom{\cdot\,,\,\cdot} &\to&
	-i\,\frac{N}{2\pi}\,\com{\cdot\,,\,\cdot}\,,\\
 \int_0^{2\pi}d\sigma^1 d\sigma^2\,* &\to&
    \frac{w_1|w_2|\Nis}{N}\int_F d\theta^1d\theta^2~\tr [\,*\,]~
	\left(\!=\frac{1}{N}\int_0^{2\pi}\!\!d\theta^1
	d\theta^2~\tr [\,*\,]\,\right) \,.
\end{eqnarray}
The first two terms of the action (\ref{eq:MLCC}) are given
by, with rescaling $\tau\to\tau/N$,
\begin{eqnarray}
 S_{2+1}&=&\frac{w_1|w_2|\Nis LT}{2}\,\int d\tau\!\int_F d\theta^1
    d\theta^2~\mathrm{Tr}\Biggl[G_{99}(F_{\tau\theta^1})^2
    +2G_{89}F_{\tau\theta^1}F_{\tau\theta^2}\nn
 &&\quad{}+G_{88}(F_{\tau\theta^2})^2-\volt(F_{\theta^1\theta^2})^2
    +(D_{\tau}X^m)^2-G_{99}(D_{\theta^1}X^m)^2\nn
 &&\quad{}-2G_{89}G_{mn}D_{\theta^1}
	X^mD_{\theta^2}X^n  -G_{88}(D_{\theta^2}X^m)^2
    +\frac{1}{2(2\pi L)^2}\,\com{X^{m},X^{n}}^2\Biggr],
	\qquad\label{D2}
\end{eqnarray}
where
\begin{eqnarray}
 F_{\tau\theta^1}&=&\ptau Y^1-\frac{R_1}{L}\,\p_{\theta^1}A
	+\frac{i}{2\pi L}\,\com{A,Y^1}\,,\\
 F_{\tau\theta^2}&=&\ptau Y^2-\frac{R_2}{L}\,\p_{\theta^2}A
	+\frac{i}{2\pi L}\,\com{A,Y^2}\,,\\
 F_{\theta^1\theta^2}&=&\frac{w_1w_2\Nis R_1R_2}{NL} I_N
    +\frac{R_1}{L}\,\p_{\theta^1}Y^2
	-\frac{R_2}{L}\,\p_{\theta^2}Y^1
	+\frac{i}{2\pi L}\,\com{Y^1,Y^2}\,,\\
 D_{\tau}X^m&=&\ptau X^m+\frac{i}{2\pi L}\,\com{A,X^m}\,,\\
 D_{\theta^1}X^m&=&\frac{R_1}{L}\,\p_{\theta_1} X^m
    +\frac{i}{2\pi L}\,\com{Y^1,X^m}\,,\\
 D_{\theta^2}X^m&=&\frac{R_2}{L}\,\p_{\theta_2} X^m
    +\frac{i}{2\pi L}\,\com{Y^2,X^m}\,.\label{DX3''}
\end{eqnarray}
As before, we rewrite the matrix regularized action by introducing
some dimensionful constants to adjust the mass dimensions of the
fields and the parameters,
\begin{eqnarray}
 Y^1(\theta^1,\theta^2)&\to&
	\hat{\alpha} A_1(x^1,x^2)\,,\label{eq:fred1}\\
 Y^2(\theta^1,\theta^2)&\to&
	\hat{\alpha} A_2(x^1,x^2)\,,\label{eq:fred2}\\
 X^m(\theta^1,\theta^2)&\to& \hat{\alpha} \phi^m(x^1,x^2)\,,\\
 A(\theta^1,\theta^2)&\to& \hat{\alpha} A_0(x^1,x^2)\,,\\
 \theta^1 &\to& x^1/\hS_1\,,\label{eq:fed5}\\
 \theta^2 &\to& x^2/\hS_2\,,\label{eq:fed6}\\
  \tau &\to& x^0/\hS\,. \label{eq:fred7}
\end{eqnarray}
where $\hat{\alpha}$ has mass dimension $-2$ and $\hS_1,
\hS_2$ and $\hS$ have mass dimension $-1$ as in
eqs.(\ref{eq:red1})-(\ref{eq:red7}).
Then, in the same way as in subsection \ref{S:MR}, in order to have
the standard form of the super Yang-Mills action, we should set
\begin{eqnarray}
 \hS&=&\frac{\hat{\alpha}}{2\pi L}\,,\label{eq:Td0'}\\
 \hS_1&=&\frac{\hat{\alpha}}{2\pi R_1}\,,\label{eq:Td1'}\\
 \hS_2&=&\frac{\hat{\alpha}}{2\pi R_2}\,.\label{eq:Td2'}
\end{eqnarray}
Thus eqs.(\ref{D2})-(\ref{DX3''}) are reduced to (we put
$\hat{\alpha}=\alpha$)
\begin{eqnarray}
 S_{2+1}&=&\frac{w_1|w_2|\Nis}{\gym^2}\int dx^0\!\int_{\calF}dx^1dx^2
    \sqrt{-\det\calG_{\alpha\beta}}~\mathrm{Tr}\Biggl[
    -\frac{1}{4}\,\calG^{\alpha\beta}
    \calG^{\gamma\delta}F_{\alpha\gamma}F_{\beta\delta}\nn
 &&{}-\frac{1}{2}\,\calG^{\alpha\beta}D_{\alpha}\phi^m
    D_{\beta}\phi^n G_{mn}+\frac{1}{4}\,G_{mp}
    G_{nq}\com{\phi^m,\phi^n}
	\com{\phi^p,\phi^q}\Biggr], \label{eq:covYMS}\\
 F_{\alpha\beta}&=&f_{\alpha\beta}+\p_\alpha A_\beta
    -\p_\beta A_\alpha+i\com{A_\alpha,A_\beta}\,,\label{Fie3''}\\
 D_\alpha\phi^m&=&\p_\alpha\phi^m+i\com{A_\alpha,\phi^m}\,,
\end{eqnarray}
where $\alpha,\beta,\gamma,\delta =0,1,2$, the worldvolume metric
$\calG_{\alpha\beta}$ is
\begin{equation}
  \calG_{\alpha\beta}= \left(\calG^{\alpha\beta}\right)^{-1},
  \quad\calG^{\alpha\beta} =\left(\begin{array}{@{\,}ccc@{\,}}
   -1& 0 & 0 \\ 0&  G_{99} & G_{89}\\
   0& G_{89} & G_{88} \end{array}\right)\,,\label{eq:WVM}
\end{equation}
the constant magnetic flux $f_{\alpha\beta}$ is
\begin{equation}
  f_{01}=f_{02}=0,\quad
  f_{12}=\frac{w_1w_2\Nis}{2\pi N\hS_1\hS_2}\,I_N\,,
\end{equation}
and the region $\calF$ in the spacesheet $(x^1,x^2)$ is a
parallelogram spanned by the two vectors
$(2\pi\hS_1s/(w_1\Nis), -2\pi\hS_2r/(w_1\Nis))$
and $(-2\pi\hS_1q/(w_2\Nis), 2\pi\hS_2p/(w_2\Nis))$.
The matrices satisfy the boundary conditions
($\Xi$ stands for $A_\alpha$ and $\phi^m$)
\begin{eqnarray}
 \Xi(x^1+\frac{2\pi\hS_1s}{w_1\Nis},
	\,x^2-\frac{2\pi \hS_2r}{w_1\Nis})
    &=&V\,\Xi(x^1,\,x^2)\,V^\dagger\,,\label{eq:BC2a}\\
 \Xi(x^1-\frac{2\pi\hS_1q}{w_2\Nis},
	\,x^2+\frac{2\pi\hS_2p}{w_2\Nis})
    &=&U\,\Xi(x^1,\,x^2)\,U^\dagger\,,\label{eq:BC2b}
\end{eqnarray}
or
\begin{eqnarray}
 \Xi(x^1+2\pi\hS_1,\,x^2)&=&
    U^{w_2r}V^{w_1p}\,\Xi(x^1,\,x^2)\,(U^{w_2r}V^{w_1p})^\dagger\,,\\
 \Xi(x^1,\,x^2+2\pi\hS_2)&=&
    U^{w_2s}V^{w_1q}\,\Xi(x^1,\,x^2)\,(U^{w_2s}V^{w_1q})^\dagger\,.
\end{eqnarray}
Note that the gauge coupling constant $\gym$ is the same as in the
flat background case in eq.(\ref{eq:GYM}).
The third term of the action (\ref{eq:MLCC}) is reduced to
\begin{eqnarray}
 S_f &=&\frac{w_1|w_2|\Nis}{2\gym^2}\int dx^0\!\!
    \int_{\calF}dx^1 dx^2\sqrt{-\det\calG_{\alpha\beta}}~\tr\Bigl[
	-iD_0\phi^l\hat{A}_{lmn}\com{\phi^m,\phi^n}\nn
 &&{}+2D_0 \phi^m(D_1\phi^n\hat{A}_{mn9}
	+D_{2}\phi^n \hat{A}_{mn8})
	-i(F_{01}\hat{A}_{mn9}
	+F_{02}\hat{A}_{mn8})\com{\phi^m,\phi^n}\nn
 &&{}+\epn^{\alpha\beta\gamma}F_{\alpha\beta}D_\gamma\phi^m
	\,\hat{A}_{m89}\Bigr]\,.\label{eq:covYMS2}
\end{eqnarray}
One comment is in order: In this subsection we have derived the matrix
regularized action by using the general matrix representation in
eqs.(\ref{eq:gr1})--(\ref{eq:gr3}). Of course, the same matrix
regularized action can be obtained when we first transform $X^9$
and $X^8$ in eqs.(\ref{eq:md1'}) and (\ref{eq:md2'}) by a $GL(2,\R)$
matrix as
\begin{equation}
  \left(\begin{array}{@{\,}c@{\,}}X^9 \\ X^8\end{array}\right)
  \to\left(\begin{array}{@{\,}c@{\,}}\tX^9 \\ \tX^8\end{array}\right)
  =\frac{1}{\Nis}\left(\begin{array}{@{\,}cc@{\,}}
    s&-rR_1/R_2\\-qR_2/R_1& p\end{array}\right)
 \left(\begin{array}{@{\,}c@{\,}}X^9 \\ X^8\end{array}\right)
\end{equation}
and then perform the regularization as in subsection \ref{S:MR}.

\subsection{Duality}
In this subsection we examine the symmetry of the matrix-regularized
action, which is the sum of $S_{2+1}$ (\ref{eq:covYMS}) and $S_f$
(\ref{eq:covYMS2}),
\begin{equation}
   \Smr=S_{2+1}+S_f\,.
\end{equation}
If we regard $x^i$ as the local coordinate of a general
two-dimensional manifold assuming $\calF\to\R^2$ or
$R_1,R_2\to0$, $\Smr$ is formally invariant under the following
two-dimensional general coordinate transformation $GC(2,\R)$,
\begin{eqnarray}
 x^i&\to&\tilde{x}^i = f^i(x)\,,\\
 \calG^{ij}(x)&\to&\tilde{\calG}^{ij}(\tilde{x})=
    M^i_{~k}M^j_{~l}\,{\calG}^{kl}(x)\,,\\
 A_i(x)&\to&\tilde{A}_i(\tilde{x})=A_j(x)(M^{-1})_{~i}^j \,,
	\label{eq:RTS}\\
 A_0(x)&\to&\tilde{A}_0(\tilde{x})=A_0(x)\,,\\
 \left(\begin{array}{@{\,}c@{\,}}
	A_{mn9}\\ A_{mn8} \end{array}\right)(x)
  &\to&\left(\begin{array}{@{\,}c@{\,}}
	\tilde{A}_{mn9}\\ \tilde{A}_{mn8}
	\end{array}\right)(\tilde{x})
   =M\left(\begin{array}{@{\,}c@{\,}}
	A_{mn9}\\ A_{mn8} \end{array}\right)(x),\\
  A_{m89}(x)&\to&\tilde{A}_{m89}(\tilde{x})
	=(\det M)\,A_{m89}(x)\,,\\
 A_{mnp}(x)&\to&\tilde{A}_{mnp}(\tilde{x})=A_{mnp}(x)\,,\\
 \phi^m(x)&\to&\tilde{\phi}^m(\tilde{x})=\phi^m(x)\,,
	\quad(i,j,k,l=1,2)
\end{eqnarray}
where
\begin{equation}
  M^i_{~j}~\left( =\frac{\p\tilde{x}^i}{\p x^j}\right)\in GL(2,\R)\,.
  \label{eq:GL2R}
\end{equation}
Note that eq.(\ref{eq:RTS}) corresponds to $GL(2,\R)$-transformation
on the (target-space) 8-9 plane in membrane theory.

Let us consider $SL(2,\R)$ transformation, which is a subgroup of the
$GC(2,\R)$,
\begin{equation}
  \tilde{\calG}^{ij}(x)
	=\Lambda^i_{~k}\,\Lambda^j_{~l}\,{\calG}^{kl}(\Lambda^{-1}x)\,,
  \quad\tilde{A}_i(x)=(\Lambda^{-1})^j_{~i}\,A_j(\Lambda^{-1}x)\,,
  \quad\tilde{\phi}^m(x)=\phi^m(\Lambda^{-1}x)\,,
	\quad\mathrm{etc.}\,,\label{eq:SL}
\end{equation}
where $\Lambda$ is a constant matrix of $SL(2,\R)$ parametrized by
\begin{equation}
  \Lambda =\left(\begin{array}{@{\,}cc@{\,}}
     a & b \\ c & d \end{array}\right)\in SL(2,\R)\,.\label{eq:SL2}
\end{equation}
It was shown that the type IIB superstring $SL(2,\R)$ duality (at the
classical level) can be realized as the $SL(2,\R)$ target-space
rotation of 11-dimensional theory in the effective action \cite{BHO}.
In particular, we can easily check that the $SL(2,\R)$ transformation
can be rewritten as (cf. Appendix \ref{S:R})
\begin{equation}
 \tilde{\tau}= \frac{c+d\tau}{a+b\tau}\,,\quad
 \tilde{\jmath}_{IJ}=|a+b\tau|\,\jmath_{IJ}\,,\quad
 \left(\begin{array}{@{\,}c@{\,}}
     \tBNS_{IJ} \\ \tBR_{IJ} \end{array}\right)
  = \Lambda \left(\begin{array}{@{\,}c@{\,}}
	\BNS_{IJ} \\ \BR_{IJ} \end{array}\right)\,,\quad
 \tilde{D}_{mnp8}= D_{mnp8}\,,\label{eq:sl2d}\quad\mathrm{etc.},
\end{equation}
where $\tau\equiv l+ie^{-\varphi}$ is the moduli fields of a 2-torus
and $I,J=1,\cdots,8$.\footnote{In this section, we put the spacetime
metric in a block diagonal form eq.(\ref{eq:blockG}) to obtain the
standard form of Yang-Mills action eqs.(\ref{eq:covYMS}) and
(\ref{eq:covYMS2}) through the matrix regularization of wrapped
membrane. Then, eq.(\ref{eq:blockG}) leads to
$B_{m8}^{(1)}=B_{m8}^{(2)}=0$ in type IIB superstring variables
(cf. Appendix \ref{S:R}). However, if we do not stick to the standard
form of Yang-Mills action and carry out the matrix regularization by
keeping the off-diagonal block non-zero, we shall see that
two-dimensional general coordinate transformation $GC(2,\R)$ leads to
the third equation of (\ref{eq:sl2d}) with non-zero $B_{m8}^{(1)}$ and
$B_{m8}^{(2)}$, in general.}
Notice that  this transformation is, in fact, corresponds to the type
IIB superstring $SL(2,\R)$ duality.
For example, we can see that when $a=d=0,\,b=-c=-1$ and $l=0$, the
$SL(2,\R)$ transformation is reduced to
the strong-weak duality $\tilde{\varphi}=-\varphi$, or $e^\varphi\to
1/e^{\varphi}$ in the type IIB superstring theory, which will be seen
in the followings.

Next, we shall examine the type IIB string duality.
Let us consider two 2-tori of $(L_1,L_2)$ and
$(\tilde{L}_1,\tilde{L}_2)$ whose metrics are $G_{uv}$ and
$\tG_{uv}$, respectively (see (\ref{eq:blockG})).
Then, following the procedure in the previous subsection, we shall
obtain the matrix regularized action of the standard form of the super
Yang-Mills action (\ref{eq:covYMS}) in each case.
We regard that those two are to be related by a $SL(2,\R)$
transformation (\ref{eq:SL})--(\ref{eq:SL2}).
Then, once we consider the reduction to type IIB superstring with each
2-torus, we shall put $R_1=R_2\equiv R_B$ and
$\tR_1=\tR_2\equiv\tR_B$ as in eq.(\ref{eq:RB})
and hence we find that the string couplings are related by
\begin{equation}
 \tgIIB =\frac{\tilde{L}_1}{\tilde{L}_2}
  =\frac{|\tG_{99}|}{\sqrt{\det\tG_{uv}}}
  =|a+b\tau|^2\frac{|G_{99}|}{\sqrt{\det G_{uv}}}
  =|a+b\tau|^2\,\frac{L_1}{L_2}
  =|a+b\tau|^2\,\gIIB\,.\label{eq:tg2b}
\end{equation}
Furthermore, since both $R_1$ and $\tR_1$ correspond to $\ell_{11}$
(cf. footnote \ref{fn:1}),
the oscillation parts of the matrices ($A_\alpha$ and
$\phi^m$) in the action are
related by the $SL(2,\R)$ transformation as (cf. (\ref{eq:Xipq}),
(\ref{eq:Ks}), (\ref{eq:fed5}) and (\ref{eq:fed6}))
\begin{eqnarray}
 \exp\Biggl[\frac{i}{N}\,\frac{K_1x^1-K_2x^2}{\hS_R}\Biggr]
  &\leftrightarrow&
    \exp\Biggl[\frac{i}{N}\,\frac{K_1y^1-K_2y^2}{\thS_R}\Biggr]
	\quad( y^i=(\Lambda^{-1})^i_{~j}\,x^j )\nn
  &&=\exp\Biggl[\frac{i}{N}\,\frac{(dK_1+cK_2)x^1-(bK_1+aK_2)x^2}
	{\hS_R}\Biggr],\quad\label{eq:tOs}
\end{eqnarray}
where $\hS_R=\alpha/(2\pi\tR_R)$ and $\thS_R=\alpha/(2\pi\tR_R)$.
In eq.(\ref{eq:tOs}) $R_B=\tR_B$ has been used and hence
$\hS_1=\hS_2\equiv\hS_R=\thS_1=\thS_2$.
We may rewrite (\ref{eq:tOs}) as the transformation of $K_i$'s,
\begin{eqnarray}
  K_1&\to&\tilde{K}_1=dK_1+cK_2\,,\label{eq:tK1}\\
  K_2&\to&\tilde{K}_2=bK_1+aK_2\,.\label{eq:tK2}
\end{eqnarray}
Since we expect $\tilde{K}_i\in\Z$, we shall put restrictions on the
parameters,
\begin{equation}
  a, b, c, d \in \Z \leftrightarrow \Lambda\in SL(2,\Z)
\end{equation}
Eqs.(\ref{eq:tK1})-(\ref{eq:tK2}) can be rewritten as the
transformation of $(p,q,r,s)$
\begin{eqnarray}
  (\,p\,~q\,)&\to& (\,\tilde{p}\,~\tilde{q}\,)=(\,p\,~q\,)
	\left(\begin{array}{@{\,}cc@{\,}}d&-b\\-c&a\end{array}\right)
	=(\,p\,~q\,)~\Lambda^{-1}\,,\label{eq:tpq}\\
  (\,r\,~s\,)&\to&(\,\tilde{r}\,~\tilde{s}\,)
	=(\,r\,~s\,)~\Lambda^{-1}\,.
\end{eqnarray}
This means that the matrices $\Xi_{(\bm{p})}(\Lambda^{-1}x)$ can be
written by $\Xi_{(\tilde{\bm{p}})}(x)$, where we have added the
suffices in order to distinguish the parameters $(p,q,r,s)$ in the
matrices (cf. (\ref{eq:Xipq})-(\ref{eq:Ks})).
Here, we should make a comment on the double dimensional reduction.
{}From eqs.(\ref{eq:Opq})-(\ref{eq:so2mx}),
(\ref{eq:fred1})-(\ref{eq:fred2}) and
(\ref{eq:Xipq})-(\ref{eq:Ks}) we shall see that the conditions of
the double dimensional reduction which corresponds to
eqs.(\ref{eq:DDR1})-(\ref{eq:DDR2}) are given by (putting $R_1=R_2$)
\begin{equation}
 pA_1(x)+qA_2(x)=0\,,\quad(q\p_1-p\p_2)\Phi(x)=0\,,\label{eq:DDR3}
\end{equation}
where $\Phi$ stands for all the matrices and background fields.
Similarly, in the $SL(2,\R)$-transformed frame the double dimensional
reduction should be done by (with $\tR_1=\tR_2$)
\begin{equation}
 \tilde{p}\tA_1(x)+\tilde{q}\tA_2(x)=0\,,\quad
  (\tilde{q}\p_1-\tilde{p}\p_2)\tilde{\Phi}(x)=0\,.\label{eq:DDR4}
\end{equation}

As was mentioned before, if we choose a $SL(2,\R)$-matrix
\begin{equation}
  \Lambda =\left(\begin{array}{@{\,}cc@{\,}}
     0 & -1 \\ 1 & 0 \end{array}\right)\in SL(2,\R)\,,
\end{equation}
eq.(\ref{eq:tg2b}) becomes
\begin{equation}
  \gIIB \to \tgIIB  =|\tau|^2\,\gIIB = l^2\gIIB+\gIIB^{-1}\,,
\end{equation}
and eq.(\ref{eq:tpq}) leads to
\begin{equation}
  (\,p\,~q\,)\to (\,\tilde{p}\,~\tilde{q}\,)=(\,-q\,~p\,)\,.
\end{equation}
This indicates that in type IIB superstring the system of a
($p,q$)-string with the string coupling $\gIIB$ is dual to that of a
($-q,p$)-string with $\gIIB^{-1}+l^2\gIIB$.
This can be seen through the terms $L_B(x)\equiv 2D_0
\phi^m(D_1\phi^n A_{mn9}+D_{2}\phi^n A_{mn8})$ in $S_f$
(\ref{eq:covYMS2}).
In the $SL(2,\R)$-transformed frame, $L_B$ is given by
\begin{eqnarray}
 \tilde{L}_B(x)&=&2\tilde{D}_0\tilde{\phi}^m(x)
    (\tilde{D}_1\tilde{\phi}^n(x)\tilde{A}_{mn9}(x)
    +\tilde{D}_{2}\tilde{\phi}^n(x)\tilde{A}_{mn8}(x))\nn
 &=&2\tilde{D}_0\tilde{\phi}^m\Bigl\{\,(\,
    \tilde{D}_1\tilde{\phi}^n\,~\tilde{D}_{2}\tilde{\phi}^n\,)
    \Lambda_{\tilde{p}\tilde{q}}^{-1}\,\Lambda_{\tilde{p}\tilde{q}}
    \left(\begin{array}{@{\,}c@{\,}}
      \tilde{A}_{mn9}(x)\\
      \tilde{A}_{mn8}(x)\end{array}\right)\,\Bigr\}\nn
 &\to&\frac{1}{\Nis}2\tilde{D}_0\tilde{\phi}^m\Bigl\{\,(\,
    (\tilde{s}\p_1-\tilde{r}\p_2)\tilde{\phi}^n\,~0\,)
    \left(\begin{array}{@{\,}c@{\,}}
	\tilde{p}\tBNS_{mn}+\tilde{q}\tBR_{mn}\\
	\tilde{r}\tBNS_{mn}+\tilde{s}\tBR_{mn}
	\end{array}\right)\,\Bigr\}\nn
 &=&\frac{2}{\Nis}\,\p_0\tilde{\phi}^m
    (\tilde{s}\p_1-\tilde{r}\p_2)\tilde{\phi}^n\,
	(\tilde{p}\tBNS_{mn}+\tilde{q}\tBR_{mn})\,.
\end{eqnarray}
One comment is in order: For a constant $GL(2,\R)$ matrix
$M=(M^i_{~j})$ in eq.(\ref{eq:GL2R}), the maps corresponding to
eqs.(\ref{eq:tg2b}) and (\ref{eq:tpq}) are given by
\begin{eqnarray}
  \gIIB&\to&\tgIIB
	=\frac{|M^1_{~1}+M^1_{~2}\,\tau|^2}{|\det M|}\,\gIIB\,,\\
  (\,p\,~q\,)&\to& (\,\tilde{p}\,~\tilde{q}\,)
	=(\,p\,~q\,)~M^{-1}\,.
\end{eqnarray}

Finally in this subsection, we refer to two specific transformations,
which are not the elements in the $SL(2,\R)$ subgroup.
First we examine the $X^8$-reflection
\begin{equation}
  M =\left(\begin{array}{@{\,}cc@{\,}}
     1 & 0 \\ 0 & -1 \end{array}\right)\in Z_2 \subset O(2)\,.
\end{equation}
Then the corresponding type IIB superstring duality is given by
\begin{eqnarray}
  &&\tilde{\tau}=-\bar{\tau}\,,\quad
	\tilde{\jmath}_{mn}=\jmath _{mn}\,,
	\quad\tilde{\jmath}_{8m}=-\jmath _{8m}\,,
	\quad \tilde{\jmath}_{88}=\jmath _{88}\,,\nn
 &&\left(\begin{array}{@{\,}c@{\,}}
    \tBNS_{IJ}\\ \tBR_{IJ} \end{array}\right)=
      \left(\begin{array}{@{\,}c@{\,}}
	\BNS_{IJ}\\ -\BR_{IJ} \end{array}\right)\,,\quad
	\tilde{D}_{mnp8}= D_{mnp8}\,.
\end{eqnarray}
In this case the type IIB string coupling is invariant and
($p,q$)-string is mapped to ($p,-q$)-string under the duality.
Note that $D_{8mnp}$ is invariant under the reflection of $X^8$ since
we have respected the symmetry of the membrane theory.
(See the discussion in ref.\cite{MO}.)

Similarly, we consider the 8-9 flip
\begin{equation}
  M=\left(\begin{array}{@{\,}cc@{\,}}
     0 & 1 \\  1 & 0 \end{array}\right)
	\in Z_2 \subset O(2)\,.
\end{equation}
Then the type IIB superstring duality is given by
\begin{eqnarray}
 &&\tilde{\tau}=\frac{1}{\bar{\tau}}\,,\quad
  \tilde{\jmath}_{mn}=|\tau|\,\jmath _{mn}\,,\quad
  \tilde{\jmath}_{8m}=-|\tau|\,\jmath _{8m}\,,\quad
  \tilde{\jmath}_{88}=|\tau|\,\jmath _{88}\,,\nn
 &&\left(\begin{array}{@{\,}c@{\,}}
     \tBNS_{IJ}\\ \tBR_{IJ} \end{array}\right)
    =\left(\begin{array}{@{\,}c@{\,}}
	\BR_{IJ}\\ \BNS_{IJ} \end{array}\right)\,,\quad
  \tilde{D}_{mnp8}= D_{mnp8}\,.
\end{eqnarray}
This implies the ($p,q$)-string $\leftrightarrow$ ($q,p$)-string and
(in $l=0$ case) the strong-weak $\gIIB \leftrightarrow \gIIB^{-1}$
duality.

\section{Summary and discussion}\label{S:SD}
In this paper we have studied matrix regularization of the wrapped
supermembrane compactified on a 2-torus. We have adopted the lightcone
wrapped supermembrane compactified on $T^2$ in the curved background
and the wrapping is characterized by two mutually prime integers
($p,q$).
We have followed the matrix regularization procedure \cite{UY4}
and also applied the double dimensional reduction technique
\cite{DHIS} properly to the matrix-regularized action as was done in
the continuous case \cite{OUY}.
We have succeeded in deducing explicitly the bosonic sector of the
matrix regularized ($p,q$)-string action in eq.(\ref{eq:pqac})
directly from the wrapped membrane.
A BPS saturated classical solution of the $(p,q)$-string action
(\ref{eq:pqac}) is valid irrespective of the value of the string
coupling $\gIIB$, however the valid region to treat the $(p,q)$-string
perturbatively is still obscure and is deserved to be
investigated.\footnote{Of course the (1,0)-string (F-string) is an
effective mode in a weak coupling region $\gIIB\ll1$, while the
(0,1)-string (D-string) in a strong coupling region $\gIIB\gg1$ for
$l=0$.}
We have also deduced the (2+1)-dimensional super Yang-Mills theory in
a curved background and then we have seen that it really has the
symmetries which are related to string duality \cite{HT,Wit}.

In this paper we have considered only classically the limit of
vanishing volume of the 2-torus with the wrapped supermembrane and it
is, of course, important to investigate it quantum mechanically.
In fact, quantum mechanical justification of the double dimensional
reduction was studied in refs.\cite{SY,UY}.
In those references, the Kaluza-Klein modes associated with the
$\rho$-coordinate were not removed classically, but they were
integrated in the path integral formulation of the wrapped
supermembrane theory. However, it is still in the beginnings of the
quantum mechanical study, and it deserves to be investigated further
with the results in this paper.

\appendix
\section{Notation}\label{S:N}
The target-space indices;
\begin{eqnarray}
  M,N,P,Q&=&1,2,\cdots,7,8,9\,,\\
  U,V,W &=&1,2,\cdots,7,y,z\,,\\
  I,J&=&1,2,\cdots,7,8\,,\\
  m,n,p,q&=&1,2,\cdots,7\,,\\
  u,v &=&8,9\,.
\end{eqnarray}
The worldvolume, worldsheet and spacesheet indices;
\begin{eqnarray}
 \alpha ,\beta &=&0,1,2\,,\\
 a,b &=& 0,1\,,\\
 i,j &=& 1,2\,.
\end{eqnarray}
The target-space metrics;
\begin{eqnarray}
  G_{MN}&=& \textrm{Target-space transverse metric} \,,\\
  \tG_{UV}&=& \textrm{Rotated target-space transverse metric} \,.
\end{eqnarray}
(Anti-)symmetrization r.w.t.\ indices;
\begin{eqnarray}
  A_{[\mu} B_{\nu]} &=& \frac{1}{2}\left(A_{\mu} B_{\nu}
	- A_{\nu} B_{\mu}\right)\,,\\
  A_{[\mu} B_{\nu} C_{\rho]} &=& \frac{1}{3!}
	(A_{\mu} B_{\nu}C_{\rho}
	+A_{\nu} B_{\rho}C_{\mu}+A_{\rho} B_{\mu}C_{\nu}\nn
  &&\quad - A_{\mu}B_{\rho}C_{\nu}
	-A_{\rho} B_{\nu}C_{\mu}-A_{\nu} B_{\mu}C_{\rho})\,,\\
  A_{[\mu} B_{|\nu|} C_{\rho]} &=& \frac{1}{2}
	(A_{\mu} B_{\nu}C_{\rho}-A_{\rho} B_{\nu}C_{\mu})\,,\\
  A_{\{\mu} B_{\nu\}} &=& \frac{1}{2}\left(A_{\mu} B_{\nu}
	+ A_{\nu} B_{\mu}\right)\,,\quad\mbox{etc.}
\end{eqnarray}

\section{Background fields}\label{S:R}
{}From the KK relation between 11-dimensional supergravity and type IIA
supergravity,
the transverse metric $G_{MN}$ can be written by
\begin{eqnarray}
 G_{MN}&\equiv& e^{-\frac{2}{3}\phi}
    \left(\begin{array}{@{\,}cc@{\,}}
	g_{IJ}+e^{2\phi}A_{I}A_{J}& e^{2\phi} A_{I}\\[10pt]
	e^{2\phi}  A_{J} & e^{2\phi} \end{array}\right)\nn
  &=& \left(\begin{array}{@{\,}cc@{\,}}
	\frac{1}{\sqrt{G_{99}}}\,g_{IJ}
	+\frac{1}{G_{99}}G_{I9}G_{J9}& G_{I9}\\[10pt]
	G_{J9} & G_{99} \end{array}\right)\nn
  &=& \left(\begin{array}{@{\,}ccc@{\,}}
	\frac{1}{\sqrt{G_{99}}}\,g_{mn}
	+\frac{1}{G_{99}}G_{m9}G_{n9}&
	\frac{1}{\sqrt{G_{99}}}\,g_{m8}
	+\frac{1}{G_{99}}G_{m9}G_{89}& G_{m9}\\[10pt]
	\frac{1}{\sqrt{G_{99}}}\,g_{8n}
	+\frac{1}{G_{99}}G_{89}G_{n9}&
	\frac{1}{\sqrt{G_{99}}}\,g_{88}
	+\frac{1}{G_{99}}G_{89}G_{89}& G_{89}\\[10pt]
	G_{n9}&G_{89} & G_{99}
			\end{array}\right)\,, \label{eq:KKmetric}
\end{eqnarray}
and the third-rank antisymmetric tensor $A_{MNP}$ is decomposed as
\begin{eqnarray}
 A_{MNP} &=& (A_{mnp}, A_{mn9},A_{mn8},A_{m89})\nn
     &=& (C_{mnp}, B_{mn},C_{mn8},B_{m8})\,.
\end{eqnarray}
Those fields are related to those of IIB as,
\begin{eqnarray}
  g_{mn}&=& \jmath_{mn}
    -\frac{\jmath_{8m}\jmath_{8n}
	-\BNS_{8m}\BNS_{8n}}{\jmath_{88}}\,,\\
  g_{8m}&=& \frac{\BNS_{8m}}{\jmath_{88}}\,,\\
  g_{88}&=&\frac{1}{\jmath_{88}}\,,\\
  C_{mn8}&=&\BR_{mn}
	+\frac{2\BR_{8[m}\jmath^{}_{n]8}}{\jmath_{88}}\,,\\
  C_{mnp}&=&D_{8mnp}
	+\frac{3}{2}\,\epn^{ij}B^{(i)}_{8[m}\,B^{(j)}_{np]}
	+\frac{3}{2}\,\epn^{ij}\frac{B^{(i)}_{8[m}\,
	B^{(j)}_{n|8|}\jmath^{}_{p]8}}{\jmath_{88}}\,,\\
  B_{mn}&=&\BNS_{mn}+\frac{2\BNS_{8[m}\jmath^{}_{n]8}}{\jmath_{88}}\,,\\
  B_{8m}&=& \frac{\jmath_{8m}}{\jmath_{88}}\,,\\
  A_m &=& -\BR_{8m} + l \BNS_{8m}\,,\\
  A_8 &=& l\,,\\
  \phi&=&\varphi -\frac{1}{2}\ln \jmath_{88}\,.
\end{eqnarray}
The modular field of a 2-torus is defined by $\tau\equiv
l+i\,e^{-\varphi }$ and can be rewritten as
\begin{eqnarray}
 \tau=\frac{G_{89}+i\sqrt{\volt}}{G_{99}}\,,
\end{eqnarray}
where $\volt \equiv G_{99}G_{88}-(G_{89})^2$.

On the other hand, the 8-9 rotated metric is given by
($U,V=1,\cdots,7,y,z$)
\begin{eqnarray}
 \tG_{UV} &=& G_{MN}\,\frac{\p X^M}{\p X^U}\,
	\frac{\p X^N}{\p X^V}\nn
  &=&\left(\begin{array}{@{\,}ccc@{\,}}
    \frac{1}{\sqrt{\tG_{zz}}}\,\tg_{mn}
	+\frac{1}{\tG_{zz}}\tG_{mz}\tG_{nz}&
	\frac{1}{\sqrt{\tG_{zz}}}\,\tg_{my}
	+\frac{1}{\tG_{zz}}\tG_{mz}\tG_{yz}& \tG_{mz}\\[10pt]
    \frac{1}{\sqrt{\tG_{zz}}}\,\tg_{yn}
	+\frac{1}{\tG_{zz}}\tG_{yz}\tG_{nz}&
	\frac{1}{\sqrt{\tG_{zz}}}\,\tg_{yy}
	+\frac{1}{\tG_{zz}}\tG_{yz}\tG_{yz}& \tG_{yz}\\[10pt]
    \tG_{nz}&\tG_{yz} &\tG_{zz}\end{array}\right)\,.\label{eq:KKRG}
\end{eqnarray}

When we choose an $SO(2)$ rotation as in eq.(\ref{eq:so2mx}) the
background metric on a 2-torus is given by
\begin{eqnarray}
 \tG_{zz}&=&\hq^2\,G_{88}+2\hp\hq\,G_{89}
		+\hp^2\,G_{99}\,, \\
 \tG_{yy}&=&\hp^2\,G_{88}-2\hp\hq\,G_{89}
		+\hq^2\,G_{99}\,,\\
 \tG_{yz}&=&\hp\hq\,G_{88}+(\hp^2-\hq^2)\,G_{89}
		-\hp\hq\,G_{99}\,.
\end{eqnarray}
The background metric on the 2-torus is rewritten by the fields of
type IIB superstring theory,
\begin{eqnarray}
 G_{99}&=& e^{4\varphi/3}\jmath_{88}^{-2/3}\,,\label{eq:2BM1}\\
 G_{89}&=& e^{4\varphi/3}\jmath_{88}^{-2/3}\,l\,,\\
 G_{88}&=& e^{4\varphi/3}\jmath_{88}^{-2/3}\,
	(l^2+e^{-2\varphi})\,,\label{eq:2BM3}\\
 \tG_{zz}&=&	e^{4\varphi/3}\jmath_{88}^{-2/3}\left\{(\hp+\hq l)^2
	+e^{-2\varphi}\hq^2\right\}\,,\label{eq:2zz}\\
 \tG_{yz}&=&
	 e^{4\varphi/3}\jmath_{88}^{-2/3}\left\{(\hp l-\hq)(\hq l+\hp)
	+\hp\hq e^{-2\varphi}\right\}\,,\label{eq:2yz} \\
 \tG_{yy}&=&
	e^{4\varphi/3}\jmath_{88}^{-2/3}\left\{(\hq-\hp l)^2
	+\hp^2e^{-2\varphi}\right\}\,.\label{eq:2yy}
\end{eqnarray}
Note that
\begin{equation}
 \sqrt{\frac{\tG_{zz}}{G_{99}}}=
	\sqrt{(\hp+\hq l)^2+e^{-2\varphi}\hq^2}\,. \label{eq:hpq}
\end{equation}
Furthermore,
\begin{equation}
 \tG_{m y} = \frac{1}{\sqrt{\tG_{zz}}}\,\tg_{m y}
	+\frac{1}{\tG_{zz}}\tG_{mz}\tG_{yz}\,.
\end{equation}
and hence the 9-dimensional metric is also rewritten by
\begin{eqnarray}
 \tg_{m y}=\frac{1}{\sqrt{\tG_{zz}}}\left(
	\tG_{m y}\tG_{zz}-\tG_{mz}\tG_{yz}\right)
 =\frac{\hp\,\BNS_{8m}+\hq\,\BR_{8m}}{%
	\jmath_{88}\sqrt{(\hp+\hq l)^2+e^{-2\varphi}\hq^2}}\,.
\end{eqnarray}
Furthermore, we shall calculate $\tg_{mn},\tg_{yy}$ as follows.
The equation,
\begin{equation}
  \tG_{yy}= \frac{1}{\sqrt{\tG_{zz}}}\,\tg_{yy}
	+\frac{1}{\tG_{zz}}\tG_{yz}\tG_{yz}\,,
\end{equation}
leads to
\begin{eqnarray}
 \tg_{yy}=\frac{1}{\sqrt{\tG_{zz}}}
	\left(\tG_{yy}\tG_{zz}-\tG_{yz}\tG_{yz}\right)
  =\frac{1}{\jmath_{88}\sqrt{(\hp+\hq l)^2+ e^{-2\varphi}\hq^2}}\,.
\end{eqnarray}
Similarly
\begin{equation}
 G_{mn}=\frac{1}{\sqrt{\tG_{zz}}}\,\tg_{mn}
	+\frac{1}{\tG_{zz}}\tG_{mz}\tG_{nz}\,,
\end{equation}
leads to
\begin{eqnarray}
 \tg_{mn}&=&\frac{1}{\sqrt{\tG_{zz}}}\left(
	G_{mn}\tG_{zz}-\tG_{mz}\tG_{nz}\right)\nn
 &=&\sqrt{(\hp+\hq l)^2+e^{-2\varphi}\hq^2}\,\left(\jmath_{mn}
	-\frac{\jmath_{8m}\jmath_{8n}}{\jmath_{88}}
    +\,\frac{(\hp\BNS_{8m}+\hq\BR_{8m})(\hp\BNS_{8n}
    +\hq\BR_{8n})}{\jmath_{88}\,\{(\hp+\hq l)^2
	+e^{-2\varphi}\hq^2\}}\right) \nn
 &=&\sqrt{(\hp+\hq l)^2+e^{-2\varphi}\hq^2}\,\left(\jmath_{mn}
	-\frac{\jmath_{8m}\jmath_{8n}}{\jmath_{88}}
    +\frac{B^{(pq)}_{8m}B^{(pq)}_{8n}}{\jmath_{88}}\right)\,.
\end{eqnarray}



\begin{thebibliography}{99}
\bibitem{Wit}E.~Witten,
	``String Theory Dynamics In Various Dimensions,''
	Nucl.\ Phys.\ B {\bf 443}, 85 (1995) [arXiv:hep-th/9503124].

\bibitem{Tow}P.~K.~Townsend,
	``The eleven-dimensional supermembrane revisited,''
	Phys.\ Lett.\ B {\bf 350}, 184 (1995) [arXiv:hep-th/9501068].

\bibitem{BST}E.~Bergshoeff, E.~Sezgin and P.~K.~Townsend,
	``Supermembranes And Eleven-Dimensional Supergravity,''
	Phys.\ Lett.\ B {\bf 189}, 75 (1987).

\bibitem{DHIS}M.~J.~Duff, P.~S.~Howe, T.~Inami and K.~S.~Stelle,
	``Superstrings In D = 10 From Supermembranes In D = 11,''
	Phys.\ Lett.\ B {\bf 191}, 70 (1987).

\bibitem{DHS}M.~Dine, P.~Y.~Huet and N.~Seiberg,
	``Large And Small Radius In String Theory,''
	Nucl.\ Phys.\ B {\bf 322}, 301 (1989).

\bibitem{DLP}J.~Dai, R.~G.~Leigh and J.~Polchinski,
	``New Connections Between String Theories,''
	Mod.\ Phys.\ Lett.\ A {\bf 4}, 2073 (1989).

\bibitem{Asp}P.~S.~Aspinwall,
	``Some Relationships Between Dualities in String Theory,''
	Nucl.\ Phys.\ Proc.\ Suppl. {\bf 46}, 30 (1996)
	[arXiv:hep-th/9508154].

\bibitem{Sch}J.~H.~Schwarz,
	``An SL(2,Z) Multiplet of Type IIB Superstrings,''
	 Phys.\ Lett.\ B{\bf 360},13 (1995) [arXiv:hep-th/9508143];
	``Superstring Dualities,'' Nucl.\ Phys.\ Proc.\ Suppl.
	{\bf 49} 183 (1996) [arXiv:hep-th/9509148];
	``The power of M theory,''
	Phys.\ Lett.\ B {\bf 367}, 97 (1996) [arXiv:hep-th/9510086].

\bibitem{W} E.~Witten,
	``Bound states of strings and p-branes,''
	Nucl.\ Phys.\ B {\bf 460}, 335 (1996) [arXiv:hep-th/9510135].

\bibitem{OUY}  H.~Okagawa, S.~Uehara and S.~Yamada,
	``(p,q)-String In The Wrapped Supermembrane On 2-Torus: A
	Classical Analysis Of The Bosonic Sector,''
	Phys.\ Lett.\  B {\bf 639}, 101 (2006) [arXiv:hep-th/0603203].

\bibitem{dWLN}
  B.~de Wit, M.~Luscher and H.~Nicolai,
  ``The Supermembrane Is Unstable,''
  Nucl.\ Phys.\ B {\bf 320}, 135 (1989).

\bibitem{Hop} J.~Hoppe,
 	``Quantum theory of a relativistic membrane,''
	M.I.T. Ph.D. thesis, (1982).

\bibitem{dWHN} B.~de Wit, J.~Hoppe and H.~Nicolai,
	``On The Quantum Mechanics Of Supermembranes,''
	Nucl.\ Phys.\ B {\bf 305}, 545 (1988).

\bibitem{BFSS} T.~Banks, W.~Fischler, S.~H.~Shenker and L.~Susskind,
	``M theory as a matrix model: A conjecture,''
	Phys.\ Rev.\ D {\bf 55}, 5112 (1997)
	[arXiv:hep-th/9610043].

\bibitem{Mot}L.~Motl,
	``Proposals on nonperturbative superstring interactions,''
	arXiv:hep-th/9701025.

\bibitem{BS} T.~Banks and N.~Seiberg,
	``Strings from matrices,''
	Nucl.\ Phys.\ B {\bf 497}, 41 (1997) [arXiv:hep-th/9702187].

\bibitem{DVV}R.~Dijkgraaf, E.~Verlinde and H.~Verlinde,
	``Matrix string theory,''
	Nucl.\ Phys.\ B {\bf 500}, 43 (1997) [arXiv:hep-th/9703030].

\bibitem{Tay} W.~I.~Taylor,
	``D-brane field theory on compact spaces,''
	Phys.\ Lett.\ B {\bf 394}, 283 (1997)
	[arXiv:hep-th/9611042].

\bibitem{SY}Y.~Sekino and T.~Yoneya,
	``From supermembrane to matrix string,''
	Nucl.\ Phys.\ B {\bf 619}, 22 (2001)
	[arXiv:hep-th/0108176];\\
	T.~Yoneya,
	``From Wrapped Supermembrane to M(atrix) Theory,''
	arXiv:hep-th/0210243.

\bibitem{Ced} M.~Cederwall,
	``Open and winding membranes, affine matrix theory and matrix
	string theory,''
	JHEP {\bf 0212}, 005 (2002) [arXiv:hep-th/0210152].

\bibitem{UY3} S.~Uehara and S.~Yamada,
	``Wrapped membranes, matrix string theory and an infinite
	dimensional Lie algebra,''
	JHEP {\bf 0407}, 043 (2004) [arXiv:hep-th/0402012].

\bibitem{UY4} S.~Uehara and S.~Yamada,
	``From supermembrane to super Yang-Mills theory,''
	Nucl.\ Phys.\ B {\bf 696}, 36 (2004)
	[arXiv:hep-th/0405037].

\bibitem{UY2} S.~Uehara and S.~Yamada,
	``Comments on the global constraints in light-cone string and
	membrane theories,''
	JHEP {\bf 0212}, 041 (2002) [arXiv:hep-th/0212048].

\bibitem{FZ} D.~B.~Fairlie, P.~Fletcher and C.~K.~Zachos,
	``Trigonometric Structure Constants For New Infinite
	Algebras,'' Phys.\ Lett.\ B {\bf 218}, 203 (1989);\\
	D.~B.~Fairlie and C.~K.~Zachos,
	``Infinite Dimensional Algebras, Sine Brackets And
	SU(Infinity),'' Phys.\ Lett.\ B {\bf 224}, 101 (1989).

\bibitem{GRT}O.~J.~Ganor, S.~Ramgoolam and W.~Taylor,
	``Branes, Fluxes and Duality in M(atrix)-Theory,''
	Nucl.\ Phys.\ B {\bf 492}, 191 (1997)
	[arXiv:hep-th/9611202].

\bibitem{FHRS}W.~Fischler, E.~Halyo, A.~Rajaraman and L.~Susskind,
	``The Incredible Shrinking Torus,''
	Nucl.\ Phys.\ B {\bf 501}, 409 (1997)
	[arXiv:hep-th/9703102].

\bibitem{DWPP2}
  B.~de Wit, K.~Peeters and J.~C.~Plefka,
  ``Supermembranes and supermatrix models,''
  arXiv:hep-th/9712082.
 
\bibitem{DWPP3}B.~de Wit, K.~Peeters and J.~Plefka,
	``Superspace geometry for supermembrane backgrounds,''
	Nucl.\ Phys.\ B {\bf 532}, 99 (1998)
	[arXiv:hep-th/9803209].

\bibitem{CDS}
  A.~Connes, M.~R.~Douglas and A.~S.~Schwarz,
  ``Noncommutative geometry and matrix theory: Compactification on tori,''
  JHEP {\bf 9802}, 003 (1998)
  [arXiv:hep-th/9711162].
 
\bibitem{BHO} E.~Bergshoeff, C.~M.~Hull and T.~Ortin,
	``Duality in the type II superstring effective action,''
	Nucl.\ Phys.\ B {\bf 451}, 547 (1995) [arXiv:hep-th/9504081].

\bibitem{MO} P.~Meessen and T.~Ortin,
	``An Sl(2,Z) multiplet of nine-dimensional type II
	supergravity theories,'' Nucl.\ Phys.\ B {\bf 541}, 195 (1999)
	[arXiv:hep-th/9806120].

\bibitem{HT}C.~M.~Hull and P.~K.~Townsend,
	``Unity of superstring dualities,''
	Nucl.\ Phys.\ B {\bf 438}, 109 (1995) [arXiv:hep-th/9410167].

\bibitem{UY} S.~Uehara and S.~Yamada,
	``On the strong coupling region in quantum matrix string
	theory,'' JHEP {\bf 0209}, 019 (2002) [arXiv:hep-th/0207209];
	``On the quantum matrix string,'' arXiv:hep-th/0210261.

\end{thebibliography}
\end{document}